\documentclass{article}

\usepackage[preprint]{template/neurips_2021}

\usepackage[utf8]{inputenc} 
\usepackage[T1]{fontenc}    
\usepackage{hyperref}       
\usepackage{url}            
\usepackage{booktabs}       
\usepackage{amsfonts}       
\usepackage{nicefrac}       
\usepackage{microtype}      
\usepackage{xcolor}         

\usepackage{amsfonts}
\usepackage{amsmath}
\usepackage{amssymb}
\usepackage{float}
\usepackage{pifont}
\usepackage{multirow}
\usepackage{makecell}

\usepackage{graphicx}
\graphicspath{{figures/}{pictures/}{images/}{./}} 

\usepackage{caption}
\usepackage{subcaption}
\usepackage{subfiles} 

\newcommand{\cmark}{\ding{51}}%
\newcommand{\xmark}{\ding{55}}%

\definecolor{young}{RGB}{252, 229, 35}
\definecolor{old}{RGB}{56, 25, 101}
\definecolor{male}{RGB}{26, 97, 165}
\definecolor{female}{RGB}{252, 105, 19}
\definecolor{dog}{RGB}{240, 110, 27}
\definecolor{fish}{RGB}{128, 80, 173}

\newcommand{\circdog}[1][dog]{  {\Large\textcolor{#1}{\ensuremath\bullet}}  }
\newcommand{\circfish}[1][fish]{  {\Large\textcolor{#1}{\ensuremath\bullet}}  }

\newcommand{\fixed}[1]{}

\newcommand{\rulesep}{\unskip\ \vrule\ }
\newcommand{\newsubcap}[1]{\phantomcaption%
       \caption*{\figurename~\thefigure(\thesubfigure): #1}}

\title{Comparing Deep Neural Nets with UMAP Tour}

\author{%
  Mingwei Li \\
  Department of Computer Science\\
  The University of Arizona\\
  Tucson, AZ 85721 \\
  \texttt{mwli@email.arizona.edu} \\
  \And
  Carlos Scheidegger \\
  Department of Computer Science\\
  The University of Arizona\\
  Tucson, AZ 85721 \\
  \texttt{cscheid@email.arizona.edu} \\
}

\begin{document}

\maketitle

\begin{abstract}
Neural networks should be interpretable to humans. 
In particular, there is a growing interest in concepts learned in a layer and similarity between layers. 
In this work, a tool, UMAP Tour, is built to visually inspect and compare internal behavior of real-world neural network models using well-aligned, \textit{instance-level} representations.
The method used in the visualization also implies a new similarity measure between neural network layers.
Using the visual tool and the similarity measure, we find concepts learned in state-of-the-art models and dissimilarities between them, such as GoogLeNet and ResNet. 
\end{abstract}

\section{Introduction}
\label{sect:intro}


Modern neural networks outperform humans in a number of learning tasks \cite{szegedy2015going,sun2014deep,lan2019albert}\fixed{Mingwei, find some corroborating citations}.
However, due to their over-parameterized, non-linear behavior, it is often difficult to reason about how they approach these tasks.
%
%
Often, we try to understand deep neural networks via their internal representations~\cite{goldfeld2018estimating}\fixed{another cite..}.
To compare these internal representations, different notions of similarity have been proposed \cite{raghu2017svcca, morcos2018insights, kornblith2019similarity}. 
In particular, Kornblith et al. used centered kernel alignment (CKA) to measure the similarity between representations in layers of neural network models \cite{kornblith2019similarity}. 
Here, we follow their exposition.
Given a set of $n$ input instances to the learning task, let $X \in \mathbb{R}^{n \times p_1}$ be a matrix of neuron activations for the $n$ examples in some layer of a neural network.
Let $Y \in \mathbb{R}^{n \times p_2}$ be defined in the same fashion for another layer, which can come from the same or a different network. 
For each pair of activations $X$ and $Y$ in two layers, a \textit{similarity index} defines a scalar function from the pair $(X,Y)$ to $s(X,Y) \in \mathbb{R}$, where a higher score indicates a higher similarity between the two layer representations.
Compared to other existing similarity indices, the centered kernel alignment (CKA) proposed by Kornblith et al.~\cite{kornblith2019similarity} satisfies three favorable properties: it is invariant to orthogonal transformations and isotropic scaling, but \textit{not} invariant to general invertible transformation.
In addition, Kornblith et al. showed that CKA captures common features in neural network layers, such as similarities between skip connections, as well as correspondences between layers in different neural network architectures. 

Similarity indices such as CKA summarize the difference between layers.
However, a single score is not sufficient to explain \textit{instance-level} differences between two layers, and instance-level explanations are often features of effective methods~\cite{kim2018interpretability}.
Here, we propose a method to measure and use layer similarity through orthogonal Procrustes analysis~\cite{gower2004procrustes}.
Orthogonal Procrustes problems have a close relation to CKA. 
By design, it satisfies the three invariance properties of CKA. 
In addition, it gives us a natural way to \textit{align} representations between two layers.
Through this alignment, we can provide \textit{instance-level} visual comparisons between layer representations.
We will demonstrate through examples that Procrustes analysis gives a reliable similarity measure as CKA does, and show use cases of utilizing such similarity index as well as the induced instance-level alignment to visually compare neural network models in our interactive visual system.
Our contributions are:

\begin{itemize}
\item We propose a similarity index between layers which satisfies the three favorable invariance properties of CKA~\cite{kornblith2019similarity};
\item We show that this similarity index comes with a reliable \textit{alignment} between data representations of different layers of neural network architectures;
\item We built an interactive visualization that aligns data representations in different layers or different neural architectures;
\item We demonstrate how the visualization identifies important events and patterns
a) when training neural networks;
b) when passing examples through the network layers;
c) when comparing two neural network architectures.
\item We provide a prototype for the visual system with multiple example models. \footnote{\url{https://umap-tour.github.io/}} 
\end{itemize}


\section{Background and Related Work}\label{sect:background}


\subsection{Dimensionality Reductions}\label{sect:bg-dim-reduction}
Due to its over-parametrized design, neural network processes and represents the input signal in very high dimensional space.
This high dimensional representation can be difficult to understand or visualize.
Fortunately, this representation is often overcomplicated, as most of its signal embeds in a lower dimensional manifold.
Many approaches has been used to reduce high dimensional data into low dimensional representation \cite{wold1987principal,borg2005modern,tenenbaum2000global,roweis2000nonlinear,van2008visualizing,mcinnes2018umap}.
In particular, Uniform Manifold Approximation and Projection (UMAP) \cite{mcinnes2018umap} can reliably preserve the global structure of high dimensional data in the low dimensional embedding.

\subsubsection{UMAP}\label{sect:bg-umap}
\begin{figure}[t]
\centering
\begin{subfigure}{.13\columnwidth}
  \centering
  \includegraphics[width=\linewidth]{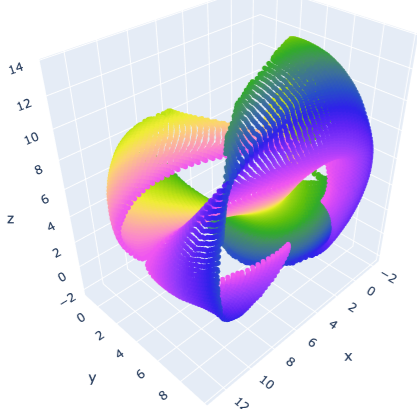}
  \caption{}
  \label{fig:klein-3d}
\end{subfigure}
\begin{subfigure}{.13\columnwidth}
  \centering
  \includegraphics[width=\linewidth]{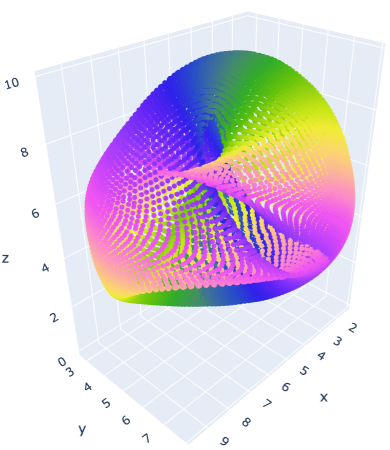}
  \caption{}
  \label{fig:klein-4d}
\end{subfigure}
\begin{subfigure}{.33\columnwidth}
  \centering
  \includegraphics[width=\linewidth]{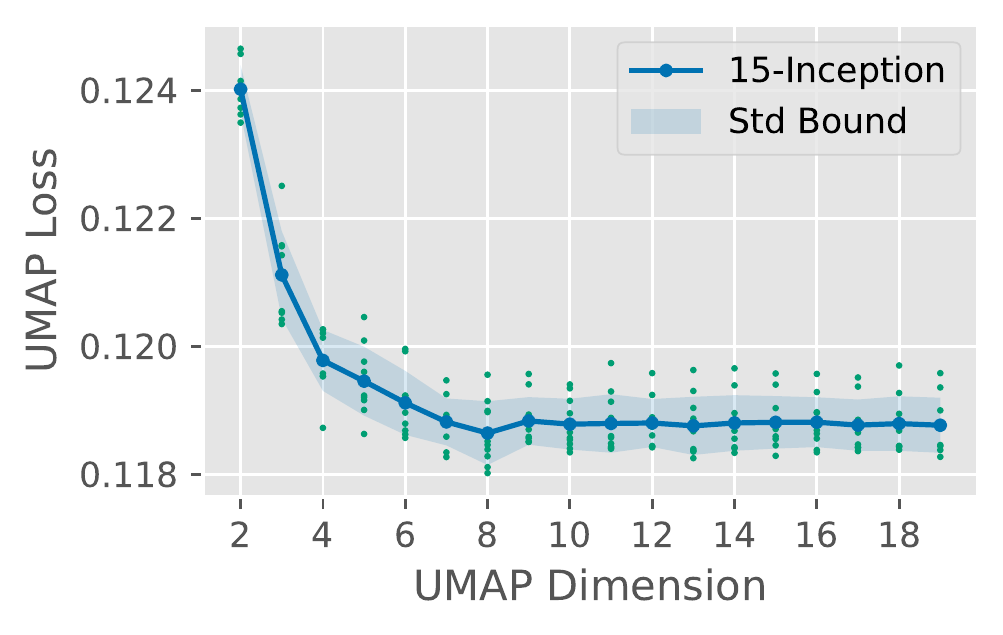}
  \caption{}
  \label{fig:umap-dim-analysis-15}
\end{subfigure}%
\begin{subfigure}{.33\columnwidth}
  \centering
  \includegraphics[width=\linewidth]{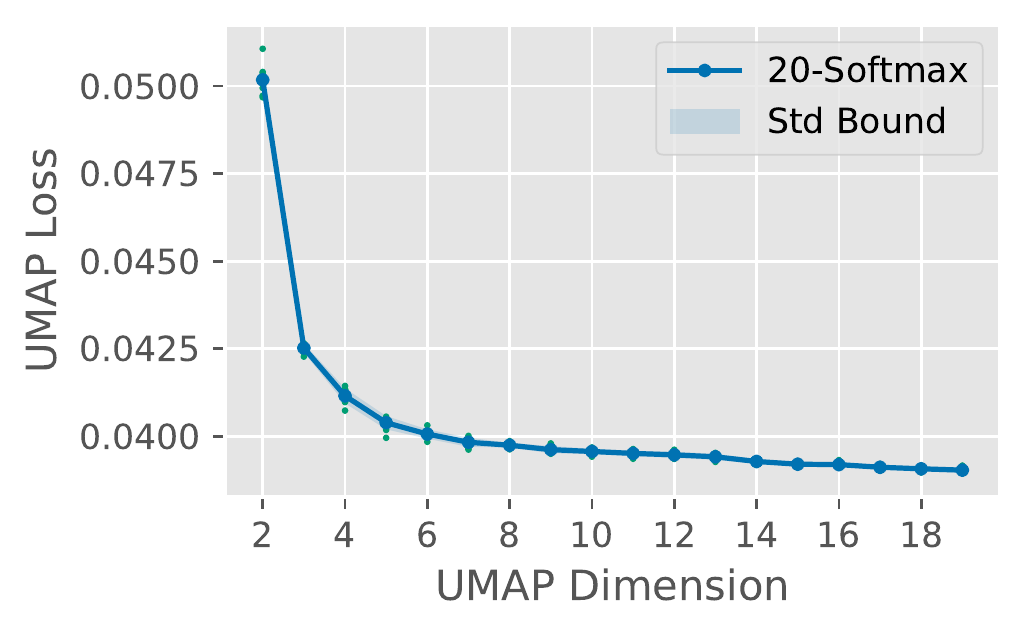}
  \caption{}
  \label{fig:umap-dim-analysis-20}
\end{subfigure}
\caption{
\textbf{(a)} Embedding a Klein bottle in 3D. The surface is tore apart due to UMAP's repulsive force.
\textbf{(b)} Embed Klein bottle in 4D and visualize in 3D. The surface is displayed continuously but has self-intersections. 
\textbf{(c) and (d)} Embedding neuron activation to more dimensions continuously improves the UMAP embedding loss.
\textbf{(c)} In the 15-Inception layer ($40768$-D), the loss decreases until 8 dimensions and stabilized afterward. 
\textbf{(d)} In 20-Softmax ($1000$-D), loss continues to improve till the largest dimension we considered.}
\label{fig:umap-dim-analysis}
\end{figure}

UMAP works by finding a low-dimensional embedding that matches the weighted neighbor graph of data points in the high dimensions.
UMAP scales well with the number of data points~\cite{mcinnes2018performance} and has little computational overhead on the embedding dimension, which make it ideal for projecting data to more than 2 or 3 dimensions.
Projecting data to more than 3 dimensions can be useful in many contexts.
For instance, one can extract low dimensional features from the original data and use the features for downstream machine learning tasks.
Here, we use multi-dimensional UMAP for visualization.
We will demonstrate that UMAP extracts important features from the very high dimensional neuron activations, and >3D embedding more reliably captures structures in the data. 
First, we will motivate our use of multi-dimensional UMAP through a toy example.
Next, we will validate this idea with real data and quantify UMAP's efficacy with increasing number of embeddings dimensions.
High dimensional data can take complex topology, which may or may not be able to reliably represented by 2 or 3 dimensional embeddings. 
For example, the space of natural image (patches) may contain structure of a Klein bottle \cite{carlsson2008local}, requiring at least $4D$ to reliably embed this structure.
In such cases, 2 or 3 dimensional UMAP can be easy to visualize, but may fail to reflect the real global data structure; 
using more than 3 dimensions can reliably preserve the data topology, meanwhile challenging to visualize.
Figure~\ref{fig:klein-3d} and \ref{fig:klein-4d} demonstrate this trade off: when embedding a Klein Bottle in 3D, the continuous surface is tore apart due to UMAP's repulsive force around its self-intersection;
when embedding the data in 4D, we can visualize data in 3D, but any single 3D projection will hallucinate a self-intersection that does not actually happen in 4D. 
This benefit and challenge of using multi-dimensional UMAP embeddings motivate our use of Grand Tour to visualize the embeddings.

On realistic data, multi-dimensional UMAP also demonstrates its strength in preserving data structures.
As we have described, UMAP works by preserving the neighbor graph. 
UMAP achieves this by finding an embedding that minimizes the binary cross entropy loss between the true neighborhood graph in the data space and the \textit{induced} neighborhood graph in the embedding.
Having more dimensions in the embedding space naturally gives UMAP more room to optimize the layout and therefore yields a lower loss.
Here we evaluate the impact of embedding dimensions on neuron activation data.
In Figure~\ref{fig:umap-dim-analysis-15} and \ref{fig:umap-dim-analysis-20}, we plot the loss of UMAP as we vary the embedding dimensions.
When applying UMAP on the activation of 15-Inception layer of a well-trained GoogLeNet model (Figure~\ref{fig:umap-dim-analysis-15}), we see a large gain from 2D to 4D, followed by smaller improvement from 4D to 8D, and a plateau afterward.
On the final softmax layer (Figure~\ref{fig:umap-dim-analysis-20}), we see the loss continues to improve until as far as 19 dimensions.
In the supplementary material we report the same analysis on other layers of GoogLeNet.
This empirical analysis of UMAP dimensionality suggests a potential benefit of embedding data to more than 3 dimensions. 
In theory, one should embed data into a space with dimensionality no less than the intrinsic dimensionality of the data manifold, as $n$-manifolds can not be homeomorphically embedded in less than $n$ dimensions.
In practice, however, it is not straightforward to estimate the intrinsic dimensionality of data manifolds and choose the optimal embedding dimension accordingly.
Throughout the experiments and visualizations in this work, we empirically use $15$ dimensional UMAP to maintain the structure in the embedding while ensuring a good real-time performance of the visualization.

\subsection{Neural Network Visualizations}\label{sect:bg-vis}
People have been visualizing neural networks via different angles, including saliency maps \cite{selvaraju2017grad,zeiler2014visualizing,bach2015pixel,ancona2019explaining}, feature visualizations \cite{olah2017feature}, embedding methods \cite{karpathy2012tsne,rauber2016visualizing,li2020visualizing}, loss landscapes\cite{li2018visualizing} or combinations of methods \cite{olah2018building,carter2019exploring}.
While most embedding-based visualizations project the high-dimensional data into 2 or 3 dimensions,
Li et al.~\cite{li2020visualizing} linearly project the internal representation onto more than 3 dimensions (using PCA) and used the Grand Tour to visualize the neuron activations in small neural networks. 
We take a similar visual approach to Li et al., but extend their method in scalability and generality.
Instead of linearly project neuron activations, we non-linearly project them via UMAP.
Our method is more flexible and suitable for larger and real-world models.
Besides, Li et al. were only able to align \textit{consecutive} layers using the singular value decomposition (SVD) of fully connected layers.
Our method aligns \textit{any} pair of layers (Section~\ref{sect:bg-procrustes}).
Our method works even when the two layers come from different neural network architectures. 
Due to this, our method can be used to \textit{compare} neural architectures (Section \ref{sect:result}).

\subsubsection{The Grand Tour}\label{sect:bg-grand-tour}
The Grand Tour \cite{asimov1985grand} is a general method to visualize high dimensional data. 
It works by rotating the multi-dimensional data points and projecting it to 2D for display. 
It is a multi-dimensional analogy to `filming a spinning 3D object and display it on a 2D screen'.
People make sense of high dimensional data through this animated 2D scatter plot rendered by a smooth sequence of 2D projections.
Formally, the Grand Tour designs a smooth sequence of orthogonal matrices $GT_t \in \mathbb{R}^{p \times p}$ (e.g. using torus method \cite{asimov1985grand}) which is parametrized by time step $t$. 
It then uses this sequence to project p-dimensional data points $X \in \mathbb{R}^{n \times p}$ and we visualize the first $2$ components of the projection. 
In other words, we visualize smoothly animated data points $\{(x_{i}(t), y_{i}(t)), i=1\dots n\}$ in a 2D scatter plot, where each data point $X_i \in \mathbb{R}^{1\times p}$ is handled by the Grand Tour rotation $GT_t$ followed by an orthogonal projection $P_2: (x_1, x_2, \dots, x_p) \mapsto (x_1, x_2)$ onto the xy-plane: 
\begin{align}
(x_{i}(t), y_{i}(t)) = X_i \cdot GT_t \cdot P_2\label{eq:grand-tour}
\end{align}

Several extensions of the Grand Tour provides controls and constraints to the projection \cite{cook1995grand,cook1997manual,cook2008grand}.
Cook et al.~\cite{cook1997manual} provides manual controls through tuning the contribution of each variable.
Instead of controlling the contribution of variables in the projection \cite{cook1997manual}, we take the more intuitive approach proposed by~\cite{li2020visualizing}.
As a result, our interface allows the viewer to directly drag a subset of data points to any desired location. 

\subsection{ Similarity Measures and Alignments}
Previous studies have proposed different notions of similarities between neural network representations \cite{raghu2017svcca,morcos2018insights,kornblith2019similarity}.
Kornblith et al.~\cite{kornblith2019similarity} observed that centered kernel alignment (CKA) has three desirable invariance properties: invariance with respect to rotations, isotropic scaling while being \textit{not} invariant to general invertible linear transforms.
In Table~\ref{table:magic}, we compare CKA, our proposal, and other methods.

\begin{table*}[t]
\caption{
Comparing similarity indices. 
Our method satisfies all three invariants of CKA \cite{kornblith2019similarity}. 
Other methods, such as linear regression, SVCCA~\cite{raghu2017svcca} and PWCCA~\cite{morcos2018insights} fail to satisfy some of these properties. 
In addition, our method provides a scalable, instance-level and well-aligned visualization.}
\label{table:magic}
\centering
\resizebox{\textwidth}{!}{%
\begin{tabular}{*8{l|}}
\toprule 
& Non-invariant to
& \multicolumn{2}{c|}{Invariant to}
& 
& \multirow{2}{*}{\makecell{Instance-level \\Visualization}}
& \\
& Linear Transform 
& Orthogonal Transform 
& Isotropic Scaling 
& Scalable
& 
& Alignment\\
\midrule
Others & \cmark/\xmark/conditional & \cmark/\xmark & \cmark/\xmark & ? & ? & ?\\
CKA                       & \cmark & \cmark & \cmark & \xmark & \xmark & ?\\
UMAP + Procrustes (\textbf{Ours})  & \cmark & \cmark & \cmark & \cmark & \cmark & \cmark \\
\bottomrule
\end{tabular}%
}
\end{table*}


\subsubsection{Centered Kernel Alignment}

A key insight from CKA is that instead of comparing multivariate features of examples, we can compare the structure of two representations through \textit{pairwise similarities} between input instances.
Although CKA works for any kernels, for simplicity, we only consider similarity using linear kernels (i.e. dot product). 
Given two layer representations $X$ and $Y$, the pairwise similarities among examples are encoded in the Gram matrices $K = X X^T$ and $L = Y Y^T$.
One can measure the similarity between these two structures by their dot product:
$
\langle K, L \rangle 
= \sum_{i,j = 1}^{n} K_{ij} L_{ij} 
= tr(XX^T YY^T)
= ||X^T Y||_F^2
$. 
Once $X$ and $Y$ are centralized, denoted by $\hat{X}$ and $\hat{Y}$, the dot product resembles the squared Frobenius norm of cross-covariance matrix between $X$ and $Y$:
$
\langle \hat{K}, \hat{L} \rangle
= ||\hat{X}^T \hat{Y}||_F^2
= ||cov(X^T, Y^T)||_F^2
$. 
CKA normalizes this dot product to make it invariant to isotropic scaling:
\begin{align}
CKA_{Linear}(X,Y) = \frac
{||\hat{X}^T\hat{Y}||_F^2}
{||\hat{X}^T\hat{X}||_F ||\hat{Y}^T\hat{Y}||_F }
\label{eq:linear-cka}
\end{align}
In the next session we will show that with a simple change of the norm, we can design another similarity measure which satisfy all three invariances while having a simpler geometric interpretation. 

\subsubsection{Orthogonal Procrustes Problems}\label{sect:bg-procrustes}
As we have explained, the Grand Tour smoothly rotates data using a sequence of orthogonal transformations. 
In other words, the \textit{visualization} is invariant to rotations.
Trying to align two rotationally invariant visualizations naturally gives us an orthogonal Procrustes problem~\cite{gower2004procrustes}.
The alignment also induces a new similarity index, similar to CKA, that is invariant to rotation but not invariant to other invertible linear transformations.

We first define orthogonal Procrustes problems.
Given two representations $X \in \mathbb{R}^{n \times p_1}$ and $Y \in \mathbb{R}^{n \times p_2}$, we further assume $p_1 = p_2 = p$.
Orthogonal Procrustes problem looks for an orthogonal transformation $Q \in \mathbb{R}^{p \times p}$ that minimizes the squared euclidean distance between $Y$ and the transformed $X$:
\begin{align*}
argmin_{Q}\; ||XQ - Y||_F^2\text{,\quad subject to }Q^T Q = I
\end{align*}
Geometrically, orthogonal Procrustes problem looks for the best rotation $Q$ that aligns $X$ with respect to $Y$, without scaling $X$.
The optimum is achieved when $Q^* = U V^T$, where $U$ and $V$ are the singular matrices of $X^TY$, i.e. $U \Sigma V^T = X^TY$. 
See \cite{gower2004procrustes} for the derivation.

\begin{figure}[t]
\centering
\begin{subfigure}{0.49\columnwidth}
  \includegraphics[scale=0.192]{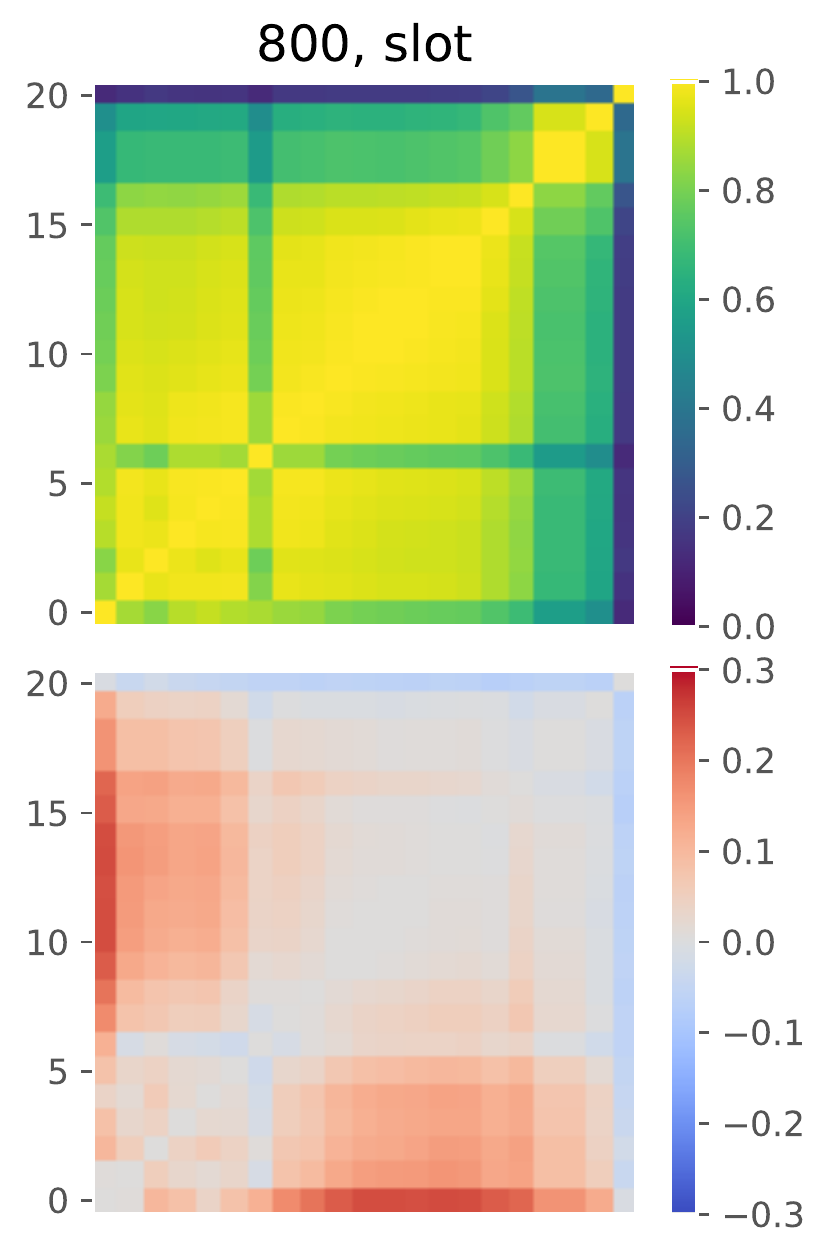}
  \includegraphics[scale=0.192]{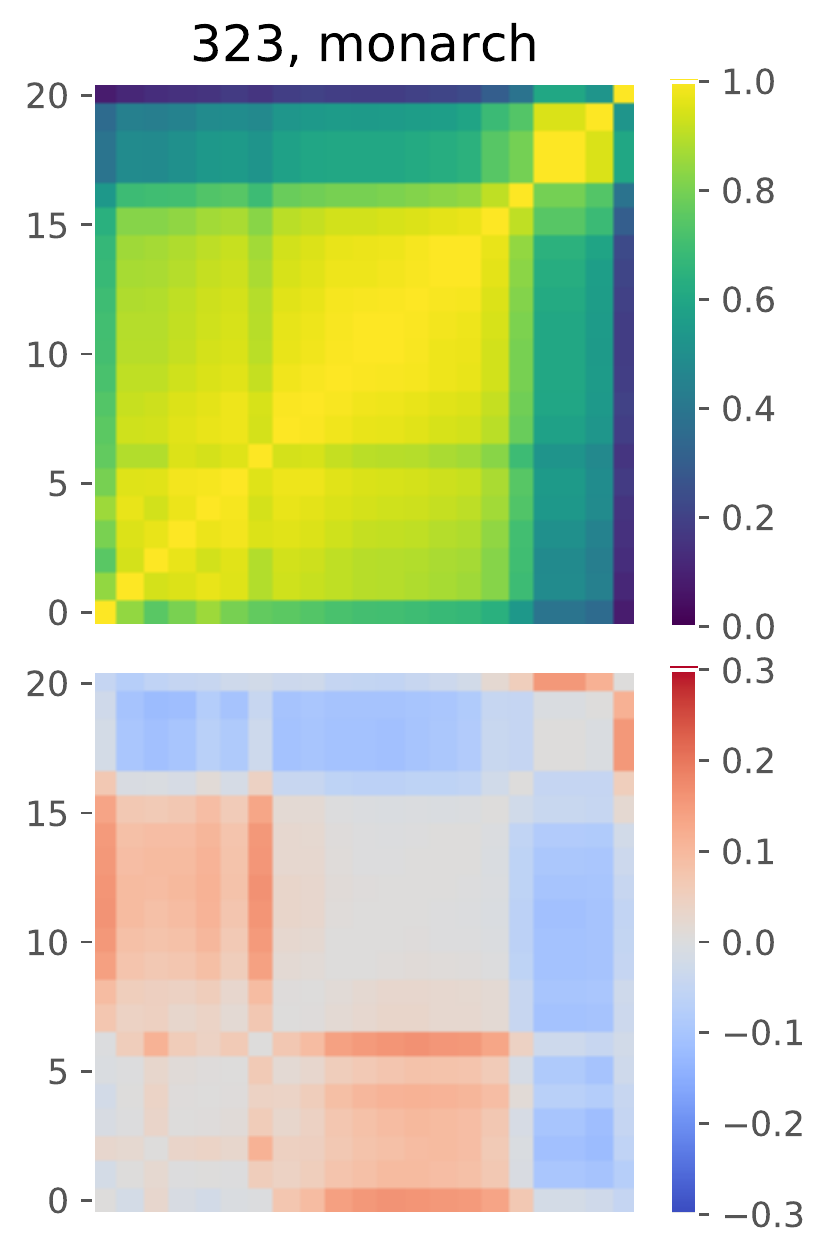}
  \includegraphics[scale=0.192]{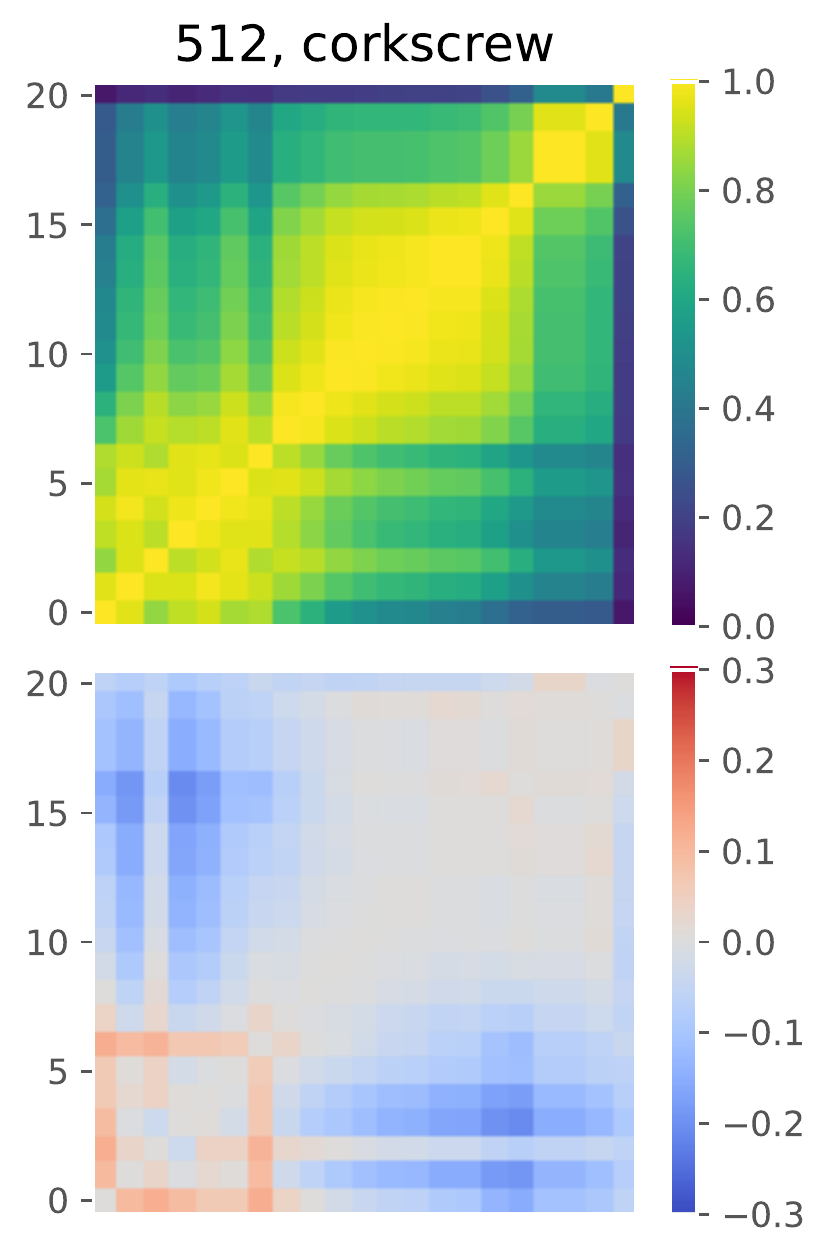}
  \includegraphics[scale=0.192]{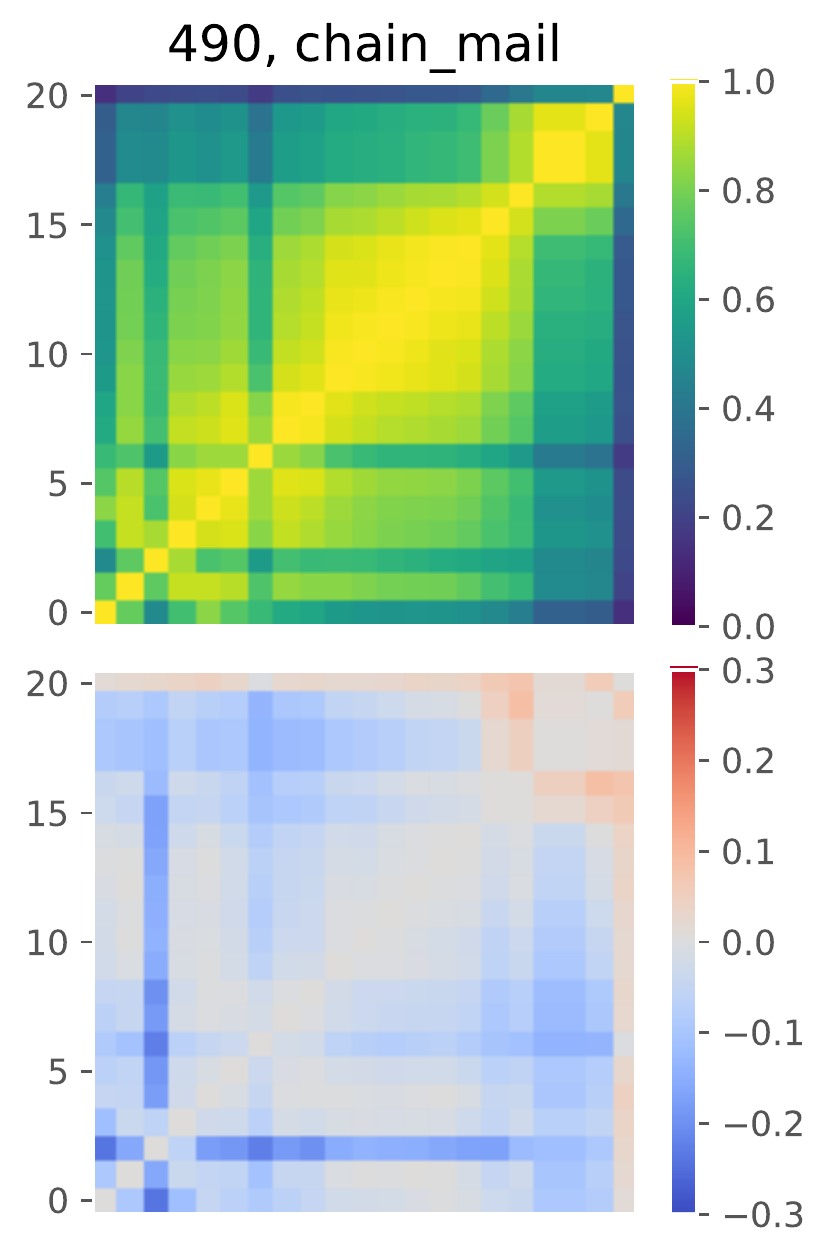}
  \newsubcap{
    CKA of GoogLeNet layers given by instances from different classes. 
    \textbf{Top:} CKA from examples from 4 individual classes. 
    \textbf{Bottom:} Deviation of each class from the mean CKA of 1000 classes.
  }
  \label{fig:cka-variance}
\end{subfigure}
\rulesep
\hfill
\begin{subfigure}{0.49\columnwidth}
  \centering
  \includegraphics[scale=0.192]{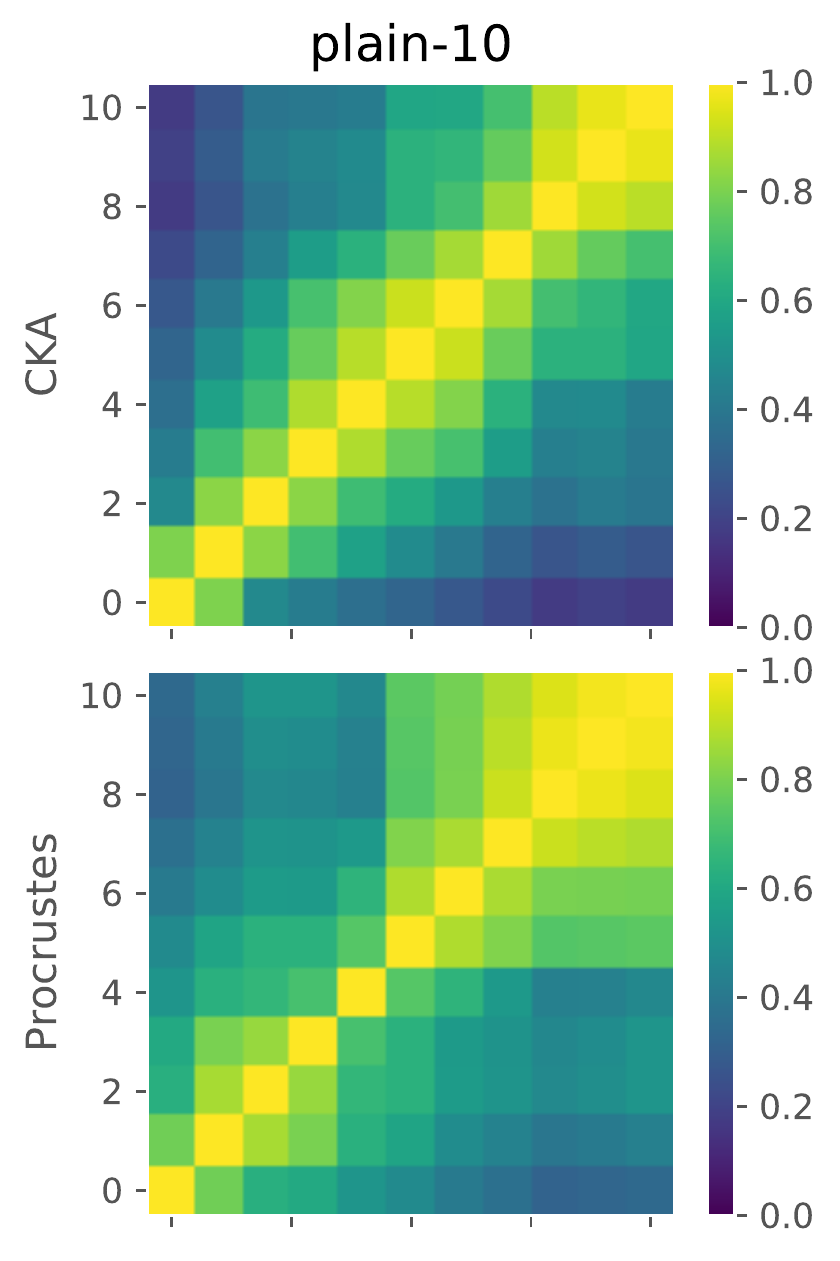}
  \includegraphics[scale=0.192]{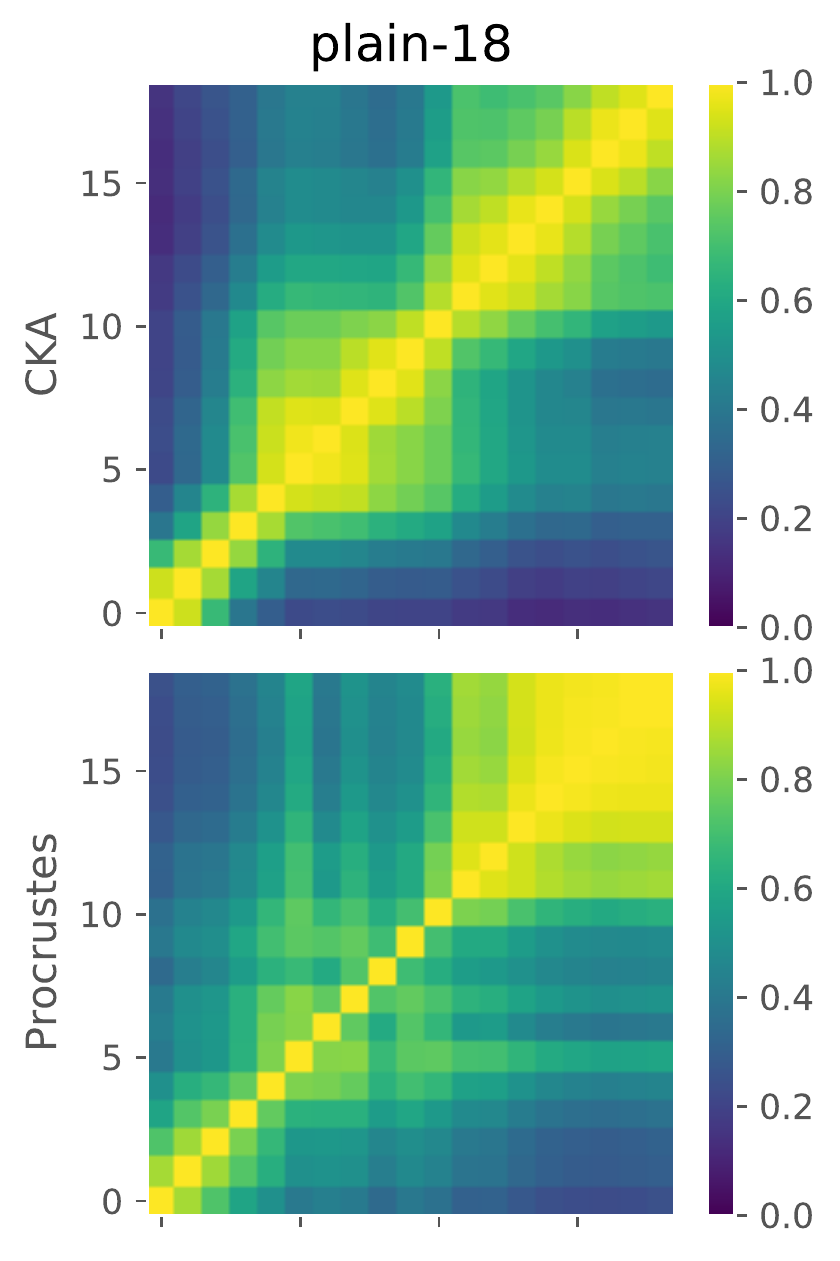}
  \includegraphics[scale=0.192]{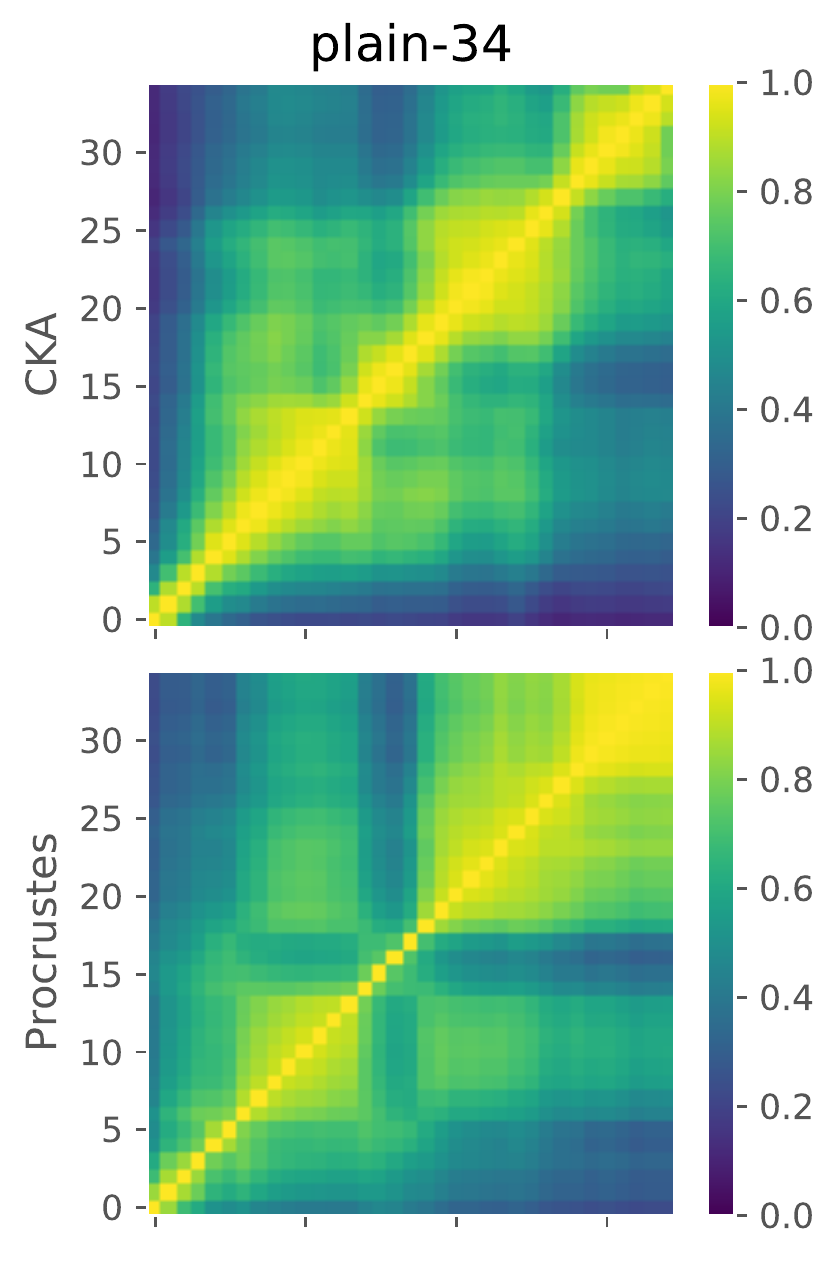}
  \includegraphics[scale=0.192]{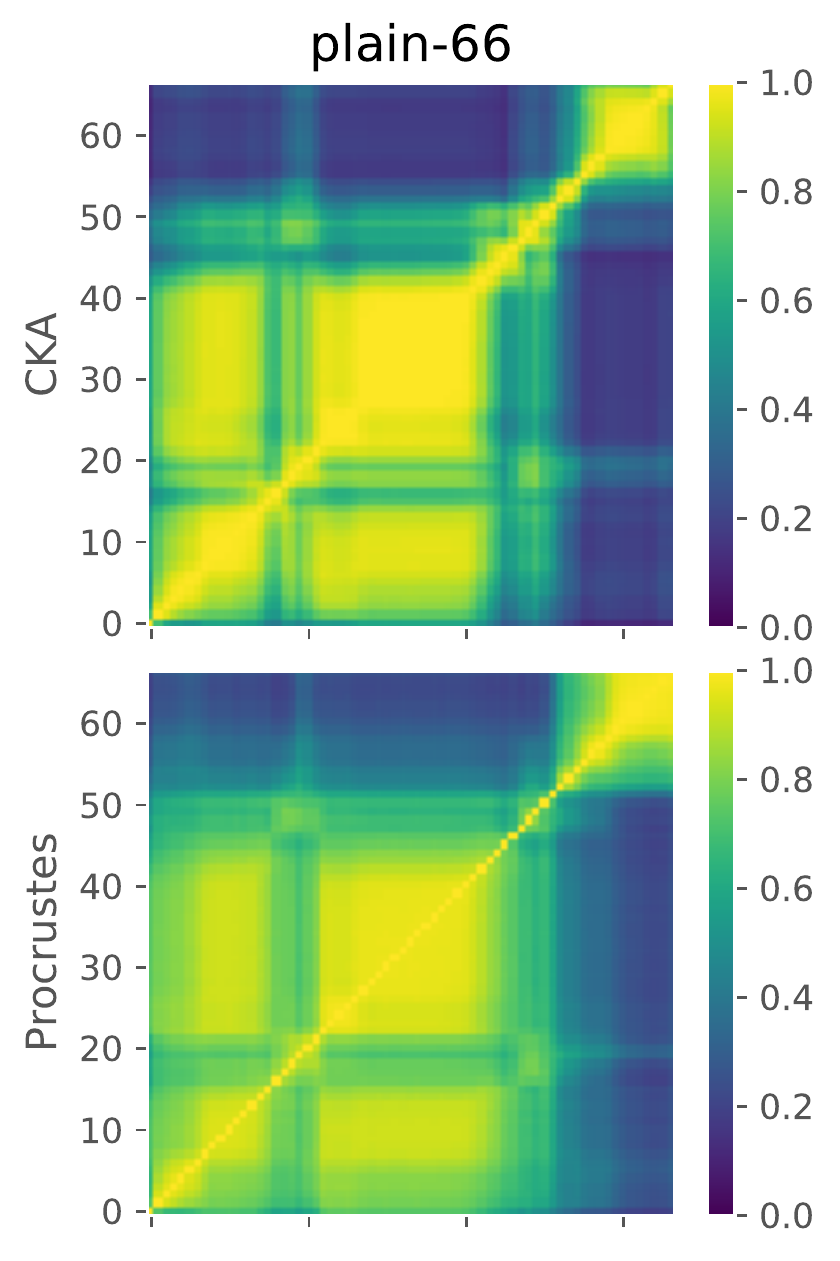}
  \newsubcap{
    Comparing linear CKA with orthogonal Procrustes. 
    Orthogonal procrustes produces similar scores and block structures to linear CKA. 
  }
  \label{fig:cka-vs-procrustes}
\end{subfigure}
\end{figure}

\textbf{Connection to CKA} 
To see how orthogonal Procrustes relates to CKA, we follow the notes in~\cite{kornblith2019similarity}. 
The squared Frobenius norm in the minimization can be expanded into
$
||XQ - Y||_F^2 
= ||X||_F^2 + ||Y||_F^2 - 2 tr(Y^T XQ)
$.
The two norms are constants, therefore minimizing $||XQ - Y||_F^2$ is equivalent to maximizing the trace $tr(Y^T XQ)$.
Plugging the optimum $Q^* = U V^T$ into the maximization shows that the maximum evaluates to 
$
max_{Q}\; tr(Y^T XQ) 
= tr(\Sigma) = ||X^TY||_*
$,  
where $||\cdot||_*$ denotes the nuclear norm, i.e. the sum of singular values of $X^TY$. 
Note this resembles the numerator of linear CKA in Eq ~\ref{eq:linear-cka}, with the squared Frobenius norm replaced by a nuclear norm.
With this connection in mind, we can define a new similarity measure based on orthogonal Procrustes. 
\begin{align}
s_{op}(X,Y) = \frac{||X^T Y||_*}{\sqrt{||X^T X||_* ||Y^T Y||_*}}
\label{eq:sim-procrsutes}
\end{align}
This is in the same form as Eq.~\ref{eq:linear-cka}, where every squared Frobenius norm is replaced by a nuclear norm. 
This similarity index satisfies all the three desired invariances that CKA possessed.
In addition, this similarity index has a straightforward geometric interpretation: it measures the degree of alignment of two configurations when rotating one of them toward the other.
More importantly, the alignment associated with this similarity can be used to match \textit{instance-level} visualizations of different layer representations. 
Here we emphasize the importance of alignment because a single number for comparing two layers is not enough.
To see why, note that similarity index depends on the choice of probing input examples. 
Probing with different examples can yield different pictures of layer-wise similarities.
Figure~\ref{fig:cka-variance} illustrates this phenomenon: when probing the layers with pictures from difference classes, the CKA reports different scores between layers.
Therefore, an instance-level visualization and easy-to-interpret alignment is needed to understand the layer behavior than a list of numbers.
Table \ref{table:magic} summarizes the strength of our method over CKA and other aligning methods discussed in \cite{kornblith2019similarity}. 
In UMAP Tour, we use orthogonal Procrustes to align two \textit{dimensionality reduced} representations.

\subsubsection{Alignment between Embeddings}\label{sect:alignment}
The UMAP embeddings on two different layers are learned independently and therefore not aligned by default.
To better compare two embeddings, we use orthogonal Procrustes to align the two embedded representations. 
We align the UMAP embeddings instead of the neuron activation vectors, to scale our method to real-world models.
Orthogonal Procrustes relies on singular value decomposition of $X^TY \in \mathbb{R}^{p \times p}$, whose runtime complexity is 
cubic to the number of features ($p$) in data \cite{golub2013matrix}.
UMAP greatly reduces the dimensionality and makes the alignment much faster in practice. 
As importantly, UMAP is able to capture most of structures in data with manageable loss of preciseness.
In Figure~\ref{fig:cka-vs-procrustes}, we compare the layer-level similarities measured by CKA on the original neuron activations with the ones measured by orthogonal Procrustes on the 15 dimensional UMAP embeddings.
We trained 4 all-convolutional nets~\cite{springenberg2014striving} with different depths on the CIFAR-10 dataset~\cite{Krizhevsky2009cifar10}, following the configuration in~\cite{kornblith2019similarity}.
Even though UMAP greatly reduced the dimensionality of data, we see the two mechanisms returns similar scores and similar block structures in the similarity matrices.
The Pearson’s r between two metrics for these four examples are 0.9493, 0.9384, 0.8953 and 0.9618 respectively, showing a consistently good linear relation between these two metrics.
%
%
%
%
\begin{figure}[t]
\centering
\includegraphics[width=0.9\columnwidth]{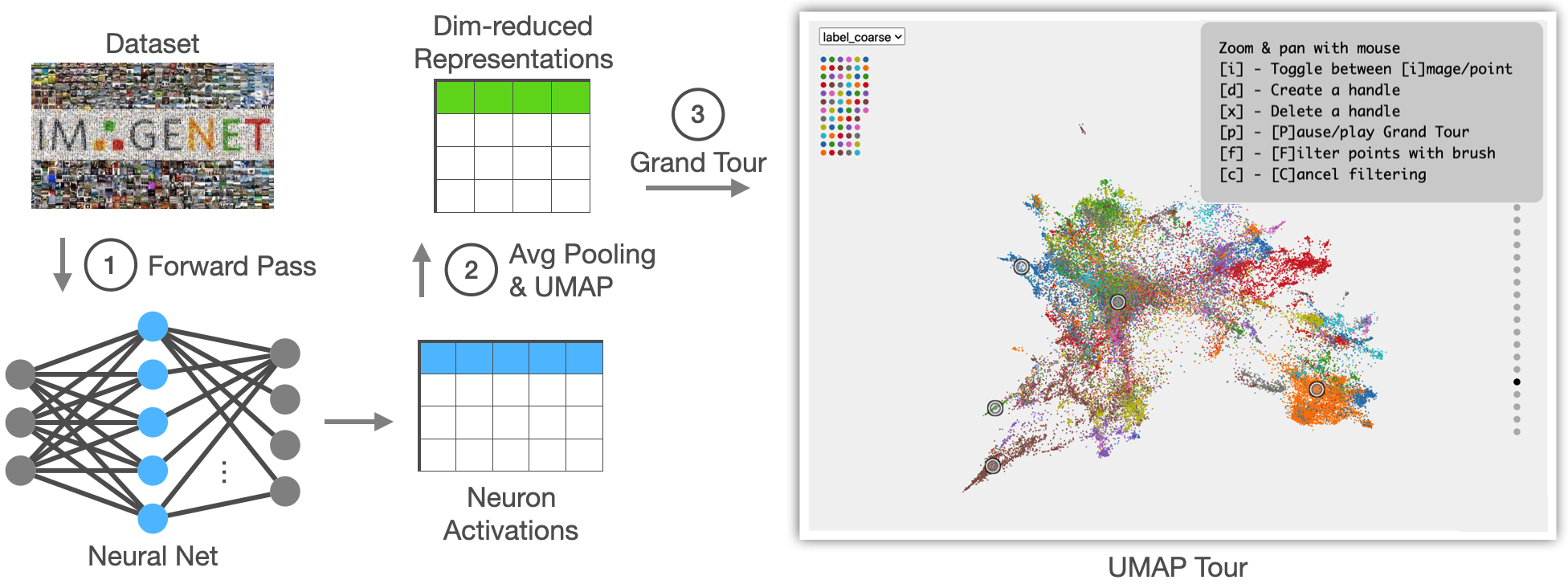}
\caption{The UMAP Tour pipeline and the user interface.}
\label{fig:pipeline}
\vspace{-12pt}
\end{figure}
%

\section{Method}\label{sect:method}

The basic steps of UMAP Tour is straightforward: we first take a dataset and pass it to the neural network of interest and record the neuron activations in every layer. 
Next, for each layer we use UMAP to project the activation to a few (e.g. $15$) dimensions. 
Finally, we use the Grand Tour to visualize the embeddings. 
The three steps are summarized in Figure \ref{fig:pipeline}. 
When moving from layer to layer or comparing two models side by side, we use procrustes transforms to align the views of instance-level representations (see Section~\ref{sect:alignment}). 


\subsection{Data Preprocessing}\label{sect:data-processing}
We start by choosing both a neural network and dataset of interest.
The dataset serves as a probe of the internal behavior of the neural network.
It does not have to be the training or validation set on which the model was trained or validated.
For example, even though GoogLeNet \cite{szegedy2015going} is most commonly trained on ImageNet \cite{deng2009imagenet}, we can either use ImageNet or other image datasets to validate concepts learned by GoogLeNet. 
In Section~\ref{sect:result}, we use a pre-trained GoogLeNet and probe it with the validation set of ImageNet \cite{deng2009imagenet}. 
In the supplementary material we include experiments with an out-of-training face dataset~\cite{karkkainen2019fairface}.
Although we are aware of the ethical issues with ImageNet, and share the concerns over its nonconsensual content~\cite{prabhu2020large}, a direct comparison to existing results in the literature requires us to use the dataset.

As the first step, we pass image examples to the network and store the neuron activations of each layer. 
The number of neurons in a layer can be very large, usually in the order of hundreds of thousand dimensions for convolutional layers, which makes it hard to visualize in practice. 
In the second step, we use UMAP to reduce the number of features to a manageable size. 
We project neuron activations to $15$ dimensions in order to capture enough interesting structures in the data. 
When fitting the UMAP embeddings, we keep all hyperparameters as default except for the embedding dimensions.~\footnote{https://umap-learn.readthedocs.io/}
We chose UMAP for its good runtime performance, flexibility and reliability to capture global structures in the data, although method works for other dimensionality reduction techniques. 
In this specific case, however, UMAP alone is still not practical for handling $\sim100$k dimensional neuron activation data. 
Even though UMAP scales reasonably well with the number of dimensions, it is still time consuming to apply UMAP directly on activations with, for example, all spatial dimensions on convolutional layers with hundreds of thousand dimensions when flattened.
To reduce the runtime of UMAP while preserving most of structural features in data, we apply average pooling before applying UMAP - we first use 2D average pooling to reduce the dimension (from hundreds of thousand) to a few tens of thousand, and then apply UMAP on the average-pooled activations. 
See supplementary for more on the computing infrastructure used, exact dimension and runtime of each layer.
%
%
%
\subsection{Visualization}\label{sect:visualization}
As the final step, we use the Grand Tour \cite{asimov1985grand} to visualize `every aspect' of the embedding space using an animated scatter plot.
At the end of the pipeline (Figure~\ref{fig:pipeline}) we show our user interface.
We used WebGL and D3.js \cite{bostock2011d3} to build a scalable web interface.
As an example, Figure~\ref{fig:pipeline} shows $50k$ images from ImageNet. 
In the central Grand Tour view, user can zoom and pan via mouse sweep.
In addition, user can steer the projection by directly selecting and dragging~\cite{li2020visualizing} data points.
On top left of the interface (Figure~\ref{fig:pipeline}), user can color data points by attributes via the option menu, such as coloring by label, prediction confidence or prediction error.
The color legend below will change accordingly.
On the right, user can switch to different layers of the model by clicking on the gray dots. 
The transitions between layers are made smooth via orthogonal Procrustes.
To compare two models, we take two Grand Tour views, align and display them side by side (Figure~\ref{fig:result-compare}, left).
We also draw the layer-wise similarities as a heatmap on the bottom. 
Clicking on any entry of the similarity matrix quickly switches to the corresponding \textit{pair} of layers. 
Other functionalities, such as direct manipulation, can be enabled via keyboard shortcuts as explained in the top right pop-up.
For example, user can switch to original images by pressing [i], manipulate projections by [d] or apply spatial filters by [f].

\subsubsection{Direct Manipulation}\label{sect:direct-manipulation}
The Grand Tour, by default, walks through all possible projections in a predetermined, but somewhat random manner. 
Controlling the projection would let users navigate to the views of their own interest, instead of waiting for them to come.
Manual control should imply changes on the entries of the Grand Tour projection matrix $GT \in \mathbb{R}^{p \times p}$.
Specifically, since data points are projected via Eq.\ref{eq:grand-tour}, the first two components in every $j^{th}$ row of the $GT$ determines where each standard basis vector $e_j$ is projected in the plot.
Before \cite{li2020visualizing}, the control is typically done via a separate interface where user selects one of the $p$ variables and change its contribution in x or y direction.
However, controlling each variable is not as intuitive as moving data points. 
This is especially true for embeddings, as each variable in the UMAP embedding may not come with semantics. 
See \cite{li2020visualizing} for details.

\subsubsection{Aligning representations}\label{sect:alignment}
\textbf{Transition between layers:} When visualizing one model with UMAP Tour, we switch the view between layers through a smooth and traceable transition. 
Specifically, when switching from one layer $Y$ to another $X$, we first align the next layer $X$ against the previous layer $Y$ via orthogonal Procrustes $Q^*$.
Next, we linearly interpolate the two aligned representations $Y$ and $XQ^*$:
$
X(s) = (1-s) \cdot Y + s \cdot XQ^*
$.
Finally, the interpolation is viewed via Grand Tour projection as in Eq.\ref{eq:grand-tour}.


\textbf{Aligning architectures:}
When visualizing two models side by side, we align their layer representations in the same manner using orthogonal Procrustes. 
When comparing the UMAP embedding of one model $Y$ with the embedding $X$ from another model, we show $Y$ and $XQ^*$ side by side in UMAP Tour. 
When directly manipulation is applied on any one of the views, the change is applied to both views so that the views are kept aligned (Figure~\ref{fig:result-compare}).

\section{Use Cases}\label{sect:result}

We test our method for visualizing a single neural network and comparing two neural networks. 
Using GoogLeNet as an example, we first visualize its training process. 
This was referred to as training dynamics by Li et al.~\cite{li2018visualizing}.
Following their terminology, we explain the layer dynamics of GoogLeNet - how examples flow through the layers. 
We found unexpected concepts learned in GoogLeNet layers, such as human faces.
Finally, we compare two different neural architectures. 
We found that they demonstrate \textit{instance-level difference} while preserving some \textit{layer-level similarity}.

\begin{figure}[t]
\centering
\includegraphics[width=\linewidth]{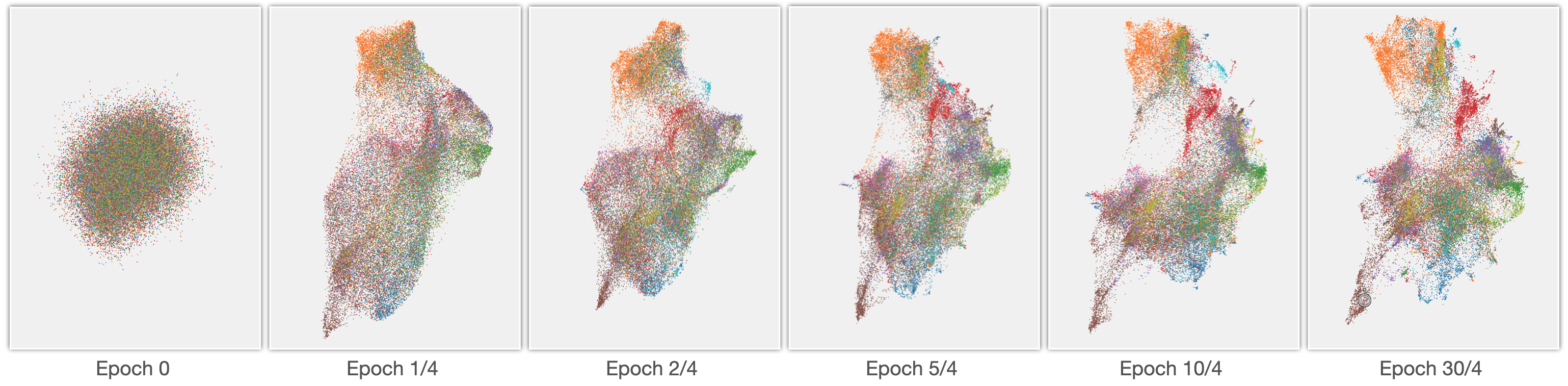}
\caption{
Training dynamics of GoogLeNet on 16-Inception layer.
}
\vspace{-12pt}
\label{fig:training-dynamics}
\end{figure}

\subsection{Training Dynamics}
We first trained a GoogLeNet from scratch with standard data augmentation strategies on the training set of ImageNet~\cite{deng2009imagenet}. 
We optimize the cross-entropy loss (on both the main and auxiliary classifiers) with an Adam~\cite{kingma2014adam} optimizer using mini-batches of size 64.
We save the state of the model once its initialized as well as on every $1/4$ epoch ($\approx 0.32M$ examples the training set). 
Figure~\ref{fig:training-dynamics} illustrates how UMAP Tour interprets a layer during this training process.
Starting from a single blob, the examples are gradually split into class-related regions throughout the training. 
Notice how dog images (orange dots on top) are first gathered on top and later spitted into smaller clusters. 

\subsection{Layer Dynamics}

Recent works~\cite{olah2017feature,cammarata2020thread,goh2021multimodal,hohman2019s} have found explanations to how neural networks learn high-level concepts from low-level features.
UMAP Tour sheds some light onto this via a different angle. 
Since an image classifier is a mapping from raw pixel colors to the label of the object, one should expect very early layers capture low-level features such as overall color and final layers summarize features into high-level concepts such as object classes.
UMAP Tour is able to show this layer dynamics and reveals how a sequence of intermediate layers approach certain goal in multiple steps.

\begin{figure}[t]
\centering
\begin{subfigure}{0.38\columnwidth}
  \includegraphics[width=\linewidth]{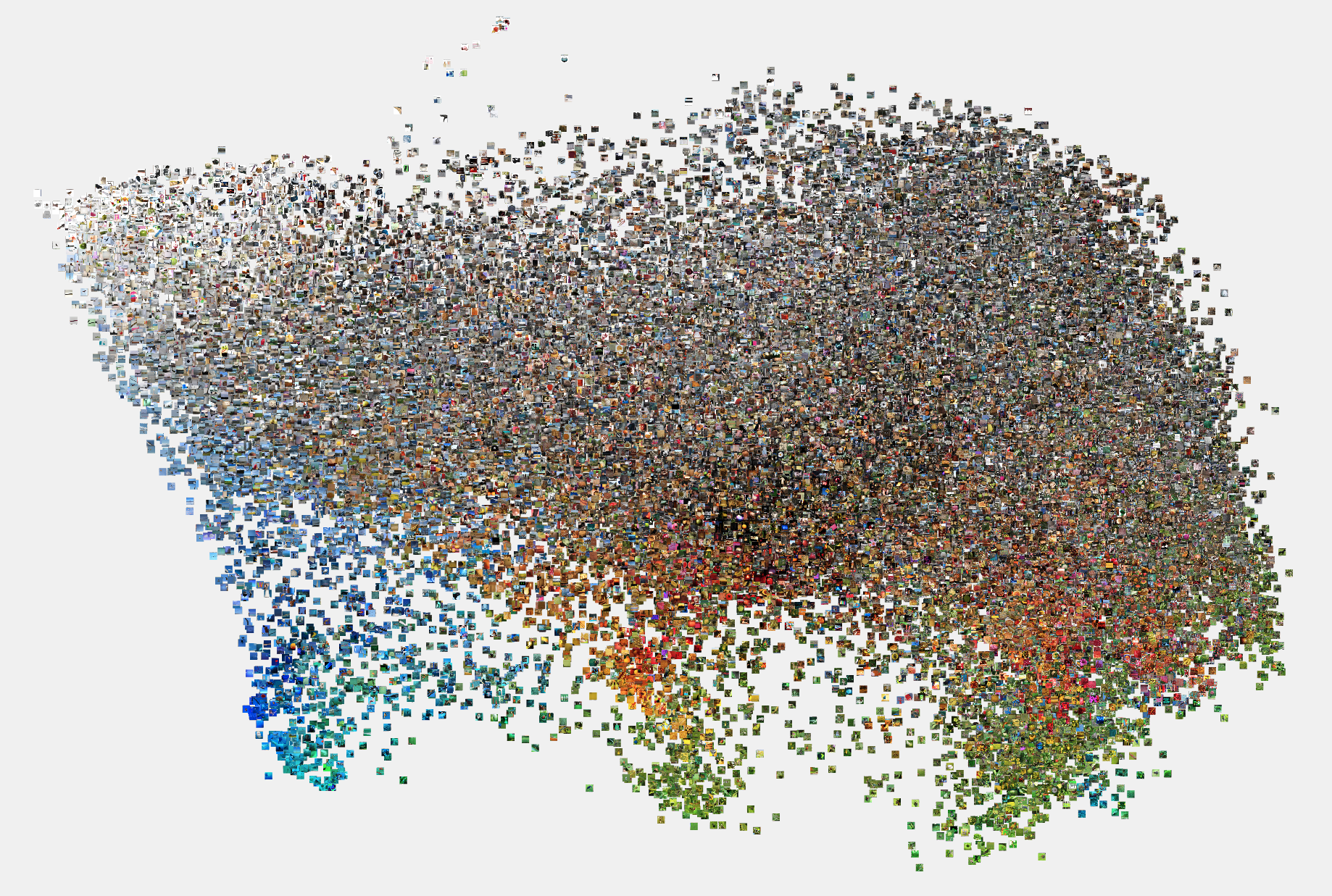}
  \newsubcap{
    Images arranged by colors in the \mbox{1-BasicConv2D} layer.
  }
  \label{fig:result-1-basicconv2d}
\end{subfigure}
\hfill
\begin{subfigure}{0.61\columnwidth}
  \includegraphics[width=\linewidth]{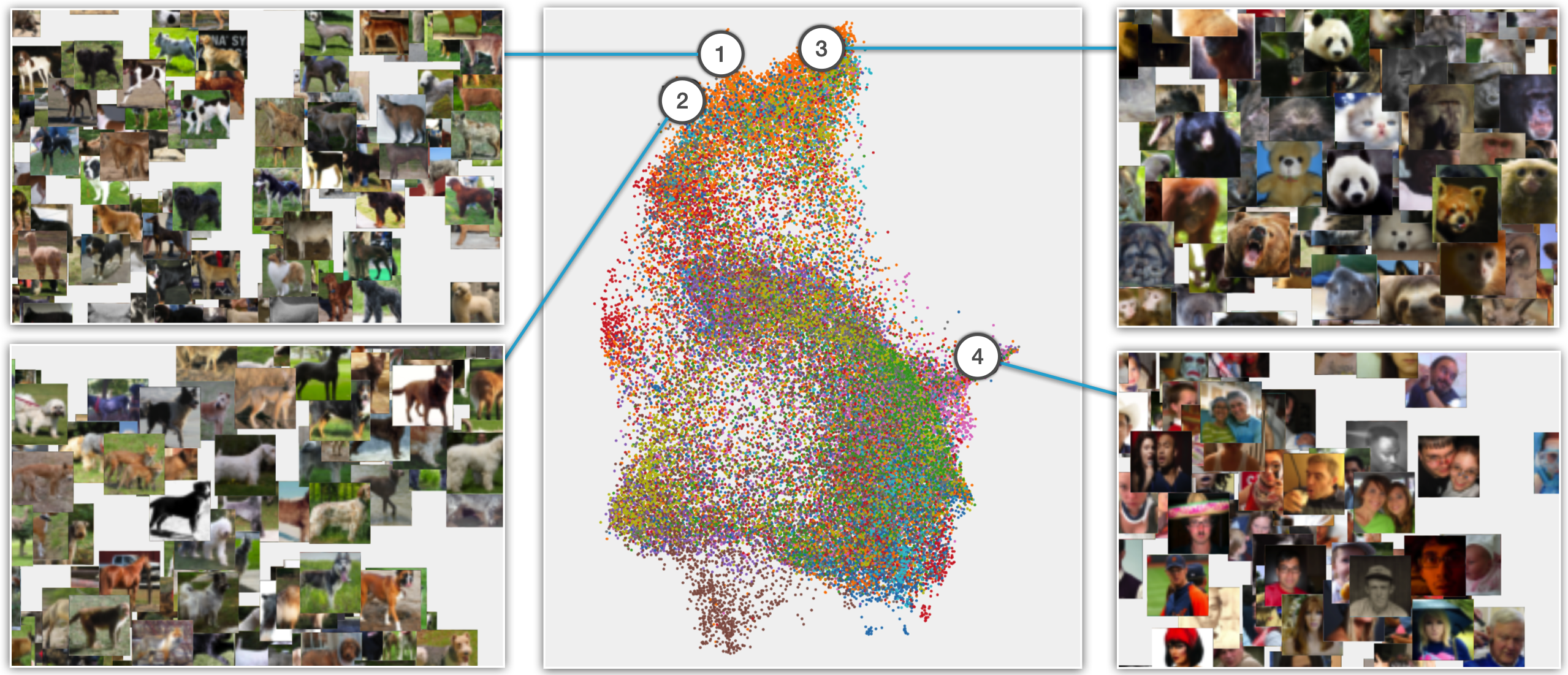}
  \newsubcap{
    Concepts learned in the 12-Inception layer: 
    \ding{192}\ding{193} Orientation of dogs; 
    \ding{194} animal faces; and
    \ding{195} human faces.
  }
  \label{fig:result-12-inception}
\end{subfigure}
\vspace{-12pt}
\end{figure}

\begin{figure*}[t]
\centering
\includegraphics[width=0.9\textwidth]{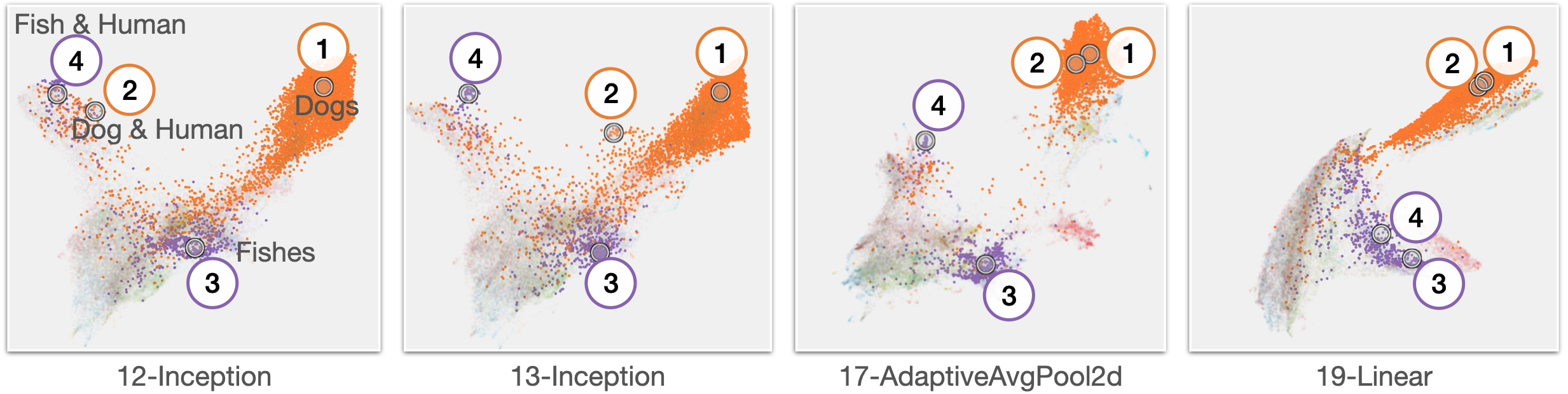}
\caption{
  Layer dynamics in GoogLeNet.
  GoogLeNet considers existence of humans as a factor when classifying other objects. 
  From the 12-Inception layer to 17-AdaptiveAvgPool2d, dog-with-human images (\ding{193}) moves from the a cluster of humans to the cluster of \circdog {\color{dog}dogs} (\ding{192}), while the fish-with-human images (\ding{195}) remain stable in the view.
  The fish-with-human (\ding{195}) moves closer to the other \circfish {\color{fish}fishes} (\ding{194}) only after the penultimate (19-Linear) layer.
}
\vspace{-10pt}
\label{fig:result-layer-dynamics}
\end{figure*}
%

At early layers, images arranged primarily by color, as shown in Figure~\ref{fig:result-1-basicconv2d}.
Later, the model encodes orientation of animals (e.g. dogs and birds) despite being only trained to classify them (Figure~\ref{fig:result-12-inception}).
In the same layer, the network also recognizes human faces, despite few human label is provided during training.
When classifying objects that may or may not appear with human (e.g. dogs alone or with their owners, fishes in the wild or held by humans), GoogLeNet considers the two cases separately (Figure~\ref{fig:result-layer-dynamics}).
Diving down from 12-Inception layer to a deeper global pooling layer (17-AdaptiveAvgPool2D), we observed a small group of dog-with-human images gradually moves toward the main cluster of dogs.
On the other hand, the cluster of images showing fish caught by human remains stable.
They only move to the cluster of fishes in the penultimate fully connected layer (19-Linear).
This demonstrates the layer dynamics of model in UMAP Tour.
The revealed patterns in each layer can be particularly useful in, for instance, finding the best layer for transfer learning.
%
%
\begin{figure}[t]
\centering
\includegraphics[width=1.0\columnwidth]{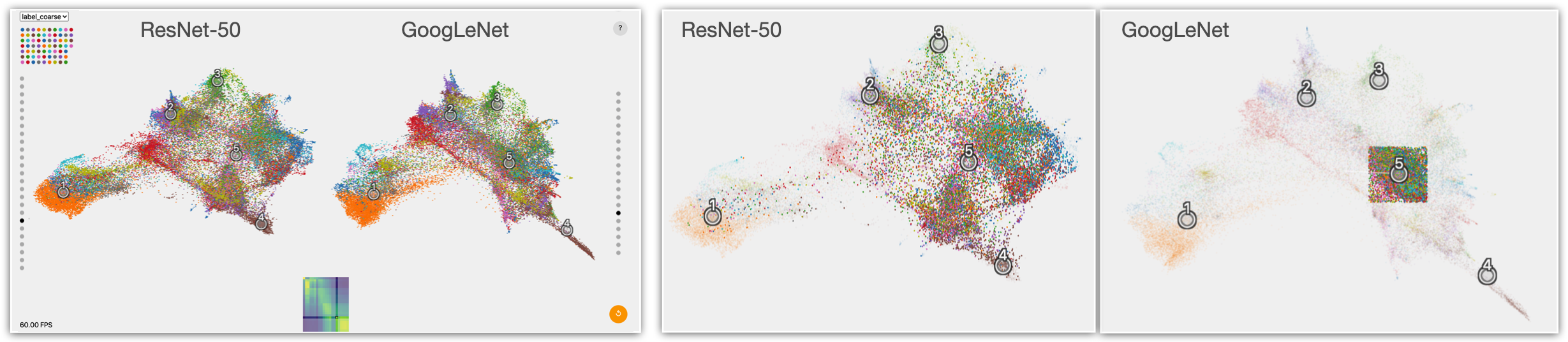}
\caption{
  Comparing ResNet-50 (18-Bottleneck layer) with GoogLeNet (15-Inception layer). 
  The two layers give a similarity score of 0.810. 
  \textbf{Left:} UI. 
  \textbf{Right:} The two views match in many landmarks, such as \ding{192} dogs, \ding{193} bugs and snakes, \ding{194} food and \ding{195} vehicles.
  However, filtering around \ding{196} reveals GoogLeNet's unique cluster of (electronics, buildings and clothings) not formed in ResNet-50.
}
\vspace{-12pt}
\label{fig:result-compare}
\end{figure}
%
\subsection{Comparing Two Models}
Although many works has been done to visually interpret the internal of neural networks \cite{olah2017feature,nguyen2019understanding, selvaraju2017grad}, fewer of them is able to directly compare two architectures.
UMAP Tour is designed for comparison especially when equipped with orthogonal Procrustes.
For example, ResNet-50 and GoogLeNet have a similar index $0.810$ on a pair of intermediate layers.
In UMAP Tour, we see similar and well-aligned UMAP embeddings in these layers, as shown by some landmarks (\ding{192}-\ding{195} in Figure~\ref{fig:result-compare}).
In addition to the similarity, UMAP Tour also shows the \textit{dissimilarity} between the two: GoogLeNet as a unique cluster of man-made objects (electronics, buildings, clothings, etc) which is more distributed in ResNet-50 (\ding{196} in Figure~\ref{fig:result-compare}).
In other words, UMAP Tour not only uses the similarity as a measure to align two representations, but also explains their \textit{dissimilarities} in its instance-level visualizations.

\section{Discussion}\label{sect:limitations}

\textbf{Orthogonal Procrustes or CKA?} 
It depends. 
The CKA can be easy to compute, since the Frobenius norm only requires summation over the square of the entries.
The similarity derived from orthogonal Procrustes, in contrast, requires finding or estimating nuclear norms.
To our knowledge, this is still under active research~\cite{heavner2019efficient}. 
In UMAP Tour, we reduce runtime by reducing data dimensionality with UMAP and believe the resultant \textit{alignment} from orthogonal Procrustes is more valuable than the similarity score.
Aligning instance-level details with orthogonal Procrustes is interpretable and intuitive - it simply rotates one configuration to another, without any scaling.
Meanwhile, although the similarity index of CKA is not invariant to general linear transform, we found its \textit{alignment} does contain scaling factors along the singular directions. 
See Appendix for details.

\textbf{Non-linear or linear dimensionality reduction?}
Although Li et al.~\cite{li2020visualizing} have advocated and demonstrated the theoretical benefits of using linear methods (i.e. PCA and the Grand Tour), we believe non-linear methods such as UMAP has its stance in practice. 
To explain a large model with millions of dimensions in one layer, PCA may require a large number of dimensions to have a good coverage of variance. 
On the other hand, rendering the Grand Tour in real time with high frame rate (e.g. 60 fps) using WebGL puts a limit to the number of dimensions we can render (e.g. by the maximum number of vertex attributes allowed in GLSL). 
To handle this trade-off, linear methods (e.g. PCA) usually put a hard stop at certain dimension while non-linear methods handle this in a more flexible way.
Within a limited budget of embedding dimensions, non-linear methods seem to summarize data relations of high dimension spaces more easily.
See Appendix for details.

\textbf{Orthogonal Procrustes or SVD?} 
Comparing to the SVD method used by Li et al.~\cite{li2020visualizing}, orthogonal Procrustes generates a slightly different alignment between linearly connected layers and works beyond linear cases.
Recall that when two layers are related by a linear transformation $Y=XW$, Li et al. align two layers by the SVD of $W$ - i.e. they align layers by $U_W V^T_W$ where $U_W$ and $V_W$ are singular matrices of $W$.
If $X$ varies uniformly along all directions (which can be the case under certain circumstances, e.g. after batch norm layers), the two methods tend to generate similar alignments, although the SVD only works for linear layers.
See Appendix for details.




\section{Conclusion}
In this work we explored how classical visual method (the Grand Tour) can be combined with modern embeddings (namely UMAP) to reason instance-level behavior of deep neural networks.
We build UMAP Tour and use it to visualize patterns within state-of-the-art models such as GoogLeNet and ResNet-50. 
The alignment method used in this visualization, namely orthogonal Procrustes on the UMAP embedding, naturally induces a new similarity measure between neural network layers.
Empirically, we find this measure numerically comparable to CKA, while the alignment gives us an intuitive, instance-level visual comparison between layers. 
Our visualization built around these notions has revealed patterns in the model internals and shed some light on the transparency and fairness concerns of neural networks.


\bibliographystyle{template/icml2020}
\newpage
\bibliography{bib}

\begin{thebibliography}{45}
\providecommand{\natexlab}[1]{#1}
\providecommand{\url}[1]{\texttt{#1}}
\expandafter\ifx\csname urlstyle\endcsname\relax
  \providecommand{\doi}[1]{doi: #1}\else
  \providecommand{\doi}{doi: \begingroup \urlstyle{rm}\Url}\fi

\bibitem[Ancona et~al.(2019)Ancona, Oztireli, and Gross]{ancona2019explaining}
Ancona, M., Oztireli, C., and Gross, M.
\newblock Explaining deep neural networks with a polynomial time algorithm for
  shapley value approximation.
\newblock In \emph{International Conference on Machine Learning}, pp.\
  272--281. PMLR, 2019.

\bibitem[Asimov(1985)]{asimov1985grand}
Asimov, D.
\newblock The grand tour: a tool for viewing multidimensional data.
\newblock \emph{SIAM journal on scientific and statistical computing},
  6\penalty0 (1):\penalty0 128--143, 1985.

\bibitem[Bach et~al.(2015)Bach, Binder, Montavon, Klauschen, M{\"u}ller, and
  Samek]{bach2015pixel}
Bach, S., Binder, A., Montavon, G., Klauschen, F., M{\"u}ller, K.-R., and
  Samek, W.
\newblock On pixel-wise explanations for non-linear classifier decisions by
  layer-wise relevance propagation.
\newblock \emph{PloS one}, 10\penalty0 (7):\penalty0 e0130140, 2015.

\bibitem[Borg \& Groenen(2005)Borg and Groenen]{borg2005modern}
Borg, I. and Groenen, P.~J.
\newblock \emph{Modern multidimensional scaling: Theory and applications}.
\newblock Springer Science \& Business Media, 2005.

\bibitem[Bostock et~al.(2011)Bostock, Ogievetsky, and Heer]{bostock2011d3}
Bostock, M., Ogievetsky, V., and Heer, J.
\newblock D3 data-driven documents.
\newblock \emph{IEEE Transactions on Visualization and Computer Graphics},
  17\penalty0 (12):\penalty0 2301–2309, December 2011.
\newblock ISSN 1077-2626.
\newblock \doi{10.1109/TVCG.2011.185}.
\newblock URL \url{https://doi.org/10.1109/TVCG.2011.185}.

\bibitem[Cammarata et~al.(2020)Cammarata, Carter, Goh, Olah, Petrov, and
  Schubert]{cammarata2020thread}
Cammarata, N., Carter, S., Goh, G., Olah, C., Petrov, M., and Schubert, L.
\newblock Thread: Circuits.
\newblock \emph{Distill}, 5\penalty0 (3):\penalty0 e24, 2020.

\bibitem[Carlsson et~al.(2008)Carlsson, Ishkhanov, De~Silva, and
  Zomorodian]{carlsson2008local}
Carlsson, G., Ishkhanov, T., De~Silva, V., and Zomorodian, A.
\newblock On the local behavior of spaces of natural images.
\newblock \emph{International journal of computer vision}, 76\penalty0
  (1):\penalty0 1--12, 2008.

\bibitem[Carter et~al.(2019)Carter, Armstrong, Schubert, Johnson, and
  Olah]{carter2019exploring}
Carter, S., Armstrong, Z., Schubert, L., Johnson, I., and Olah, C.
\newblock Exploring neural networks with activation atlases.
\newblock \emph{Distill.}, 2019.

\bibitem[Cook \& Buja(1997)Cook and Buja]{cook1997manual}
Cook, D. and Buja, A.
\newblock Manual controls for high-dimensional data projections.
\newblock \emph{Journal of computational and Graphical Statistics}, 6\penalty0
  (4):\penalty0 464--480, 1997.

\bibitem[Cook et~al.(1995)Cook, Buja, Cabrera, and Hurley]{cook1995grand}
Cook, D., Buja, A., Cabrera, J., and Hurley, C.
\newblock Grand tour and projection pursuit.
\newblock \emph{Journal of Computational and Graphical Statistics}, 4\penalty0
  (3):\penalty0 155--172, 1995.

\bibitem[Cook et~al.(2008)Cook, Buja, Lee, and Wickham]{cook2008grand}
Cook, D., Buja, A., Lee, E.-K., and Wickham, H.
\newblock Grand tours, projection pursuit guided tours, and manual controls.
\newblock In \emph{Handbook of data visualization}, pp.\  295--314. Springer,
  2008.

\bibitem[Deng et~al.(2009)Deng, Dong, Socher, Li, Li, and Li]{deng2009imagenet}
Deng, J., Dong, W., Socher, R., Li, L., Li, K., and Li, F.
\newblock Imagenet: {A} large-scale hierarchical image database.
\newblock In \emph{2009 {IEEE} Computer Society Conference on Computer Vision
  and Pattern Recognition {(CVPR} 2009), 20-25 June 2009, Miami, Florida,
  {USA}}, pp.\  248--255. {IEEE} Computer Society, 2009.
\newblock \doi{10.1109/CVPR.2009.5206848}.
\newblock URL \url{http://www.image-net.org/}.

\bibitem[Goh et~al.(2021)Goh, Cammarata, Voss, Carter, Petrov, Schubert,
  Radford, and Olah]{goh2021multimodal}
Goh, G., Cammarata, N., Voss, C., Carter, S., Petrov, M., Schubert, L.,
  Radford, A., and Olah, C.
\newblock Multimodal neurons in artificial neural networks.
\newblock \emph{Distill}, 6\penalty0 (3):\penalty0 e30, 2021.

\bibitem[Goldfeld et~al.(2018)Goldfeld, Berg, Greenewald, Melnyk, Nguyen,
  Kingsbury, and Polyanskiy]{goldfeld2018estimating}
Goldfeld, Z., Berg, E. v.~d., Greenewald, K., Melnyk, I., Nguyen, N.,
  Kingsbury, B., and Polyanskiy, Y.
\newblock Estimating information flow in deep neural networks.
\newblock \emph{arXiv preprint arXiv:1810.05728}, 2018.

\bibitem[Golub \& Van~Loan(2013)Golub and Van~Loan]{golub2013matrix}
Golub, G. and Van~Loan, C.
\newblock Matrix computations 4th edition the johns hopkins university press.
\newblock \emph{Baltimore, MD}, 2013.

\bibitem[Gower et~al.(2004)Gower, Dijksterhuis, et~al.]{gower2004procrustes}
Gower, J.~C., Dijksterhuis, G.~B., et~al.
\newblock \emph{Procrustes problems}, volume~30.
\newblock Oxford University Press on Demand, 2004.

\bibitem[Heavner \& Martinsson(2019)Heavner and
  Martinsson]{heavner2019efficient}
Heavner, N. and Martinsson, P.-G.
\newblock Efficient nuclear norm approximation via the randomized utv
  algorithm.
\newblock \emph{arXiv preprint arXiv:1903.11543}, 2019.

\bibitem[Hohman et~al.(2019)Hohman, Park, Robinson, and Chau]{hohman2019s}
Hohman, F., Park, H., Robinson, C., and Chau, D. H.~P.
\newblock S ummit: Scaling deep learning interpretability by visualizing
  activation and attribution summarizations.
\newblock \emph{IEEE transactions on visualization and computer graphics},
  26\penalty0 (1):\penalty0 1096--1106, 2019.

\bibitem[K{\"a}rkk{\"a}inen \& Joo(2019)K{\"a}rkk{\"a}inen and
  Joo]{karkkainen2019fairface}
K{\"a}rkk{\"a}inen, K. and Joo, J.
\newblock Fairface: Face attribute dataset for balanced race, gender, and age.
\newblock \emph{arXiv preprint arXiv:1908.04913}, 2019.
\newblock URL \url{https://github.com/dchen236/FairFace}.

\bibitem[Karpathy(2012)]{karpathy2012tsne}
Karpathy, A.
\newblock t-sne visualization of cnn codes, 2012.
\newblock URL \url{https://cs.stanford.edu/people/karpathy/cnnembed/}.

\bibitem[Kim et~al.(2018)Kim, Wattenberg, Gilmer, Cai, Wexler, Vi{\'{e}}gas,
  and Sayres]{kim2018interpretability}
Kim, B., Wattenberg, M., Gilmer, J., Cai, C.~J., Wexler, J., Vi{\'{e}}gas,
  F.~B., and Sayres, R.
\newblock Interpretability beyond feature attribution: Quantitative testing
  with concept activation vectors {(TCAV)}.
\newblock In \emph{Proceedings of the 35th International Conference on Machine
  Learning, {ICML} 2018, Stockholmsm{\"{a}}ssan, Stockholm, Sweden, July 10-15,
  2018}, volume~80 of \emph{Proceedings of Machine Learning Research}, pp.\
  2673--2682. {PMLR}, 2018.

\bibitem[Kingma \& Ba(2014)Kingma and Ba]{kingma2014adam}
Kingma, D.~P. and Ba, J.
\newblock Adam: A method for stochastic optimization.
\newblock \emph{arXiv preprint arXiv:1412.6980}, 2014.

\bibitem[Kornblith et~al.(2019)Kornblith, Norouzi, Lee, and
  Hinton]{kornblith2019similarity}
Kornblith, S., Norouzi, M., Lee, H., and Hinton, G.~E.
\newblock Similarity of neural network representations revisited.
\newblock In \emph{Proceedings of the 36th International Conference on Machine
  Learning, {ICML} 2019, 9-15 June 2019, Long Beach, California, {USA}},
  volume~97 of \emph{Proceedings of Machine Learning Research}, pp.\
  3519--3529. {PMLR}, 2019.

\bibitem[Krizhevsky et~al.(2009)Krizhevsky, Nair, and
  Hinton]{Krizhevsky2009cifar10}
Krizhevsky, A., Nair, V., and Hinton, G.
\newblock Cifar-10 (canadian institute for advanced research).
\newblock 2009.
\newblock URL \url{http://www.cs.toronto.edu/~kriz/cifar.html}.

\bibitem[Lan et~al.(2020)Lan, Chen, Goodman, Gimpel, Sharma, and
  Soricut]{lan2019albert}
Lan, Z., Chen, M., Goodman, S., Gimpel, K., Sharma, P., and Soricut, R.
\newblock {ALBERT:} {A} lite {BERT} for self-supervised learning of language
  representations.
\newblock In \emph{8th International Conference on Learning Representations,
  {ICLR} 2020, Addis Ababa, Ethiopia, April 26-30, 2020}. OpenReview.net, 2020.

\bibitem[Li et~al.(2018)Li, Xu, Taylor, Studer, and
  Goldstein]{li2018visualizing}
Li, H., Xu, Z., Taylor, G., Studer, C., and Goldstein, T.
\newblock Visualizing the loss landscape of neural nets.
\newblock In \emph{Proceedings of the 32nd International Conference on Neural
  Information Processing Systems}, pp.\  6391--6401, 2018.

\bibitem[Li et~al.(2020)Li, Zhao, and Scheidegger]{li2020visualizing}
Li, M., Zhao, Z., and Scheidegger, C.
\newblock Visualizing neural networks with the grand tour.
\newblock \emph{Distill}, 5\penalty0 (3):\penalty0 e25, 2020.

\bibitem[McInnes(2018)]{mcinnes2018performance}
McInnes, L.
\newblock Performance comparison of dimension reduction implementations, 2018.
\newblock URL
  \url{https://umap-learn.readthedocs.io/en/latest/benchmarking.html}.

\bibitem[McInnes et~al.(2018)McInnes, Healy, and Melville]{mcinnes2018umap}
McInnes, L., Healy, J., and Melville, J.
\newblock Umap: Uniform manifold approximation and projection for dimension
  reduction.
\newblock \emph{arXiv preprint arXiv:1802.03426}, 2018.

\bibitem[Morcos et~al.(2018)Morcos, Raghu, and Bengio]{morcos2018insights}
Morcos, A.~S., Raghu, M., and Bengio, S.
\newblock Insights on representational similarity in neural networks with
  canonical correlation.
\newblock In \emph{Advances in Neural Information Processing Systems 31: Annual
  Conference on Neural Information Processing Systems 2018, NeurIPS 2018,
  December 3-8, 2018, Montr{\'{e}}al, Canada}, pp.\  5732--5741, 2018.

\bibitem[Nguyen et~al.(2019)Nguyen, Yosinski, and
  Clune]{nguyen2019understanding}
Nguyen, A., Yosinski, J., and Clune, J.
\newblock Understanding neural networks via feature visualization: A survey.
\newblock In \emph{Explainable AI: interpreting, explaining and visualizing
  deep learning}, pp.\  55--76. Springer, 2019.

\bibitem[Olah et~al.(2017)Olah, Mordvintsev, and Schubert]{olah2017feature}
Olah, C., Mordvintsev, A., and Schubert, L.
\newblock Feature visualization.
\newblock \emph{Distill}, 2\penalty0 (11):\penalty0 e7, 2017.

\bibitem[Olah et~al.(2018)Olah, Satyanarayan, Johnson, Carter, Schubert, Ye,
  and Mordvintsev]{olah2018building}
Olah, C., Satyanarayan, A., Johnson, I., Carter, S., Schubert, L., Ye, K., and
  Mordvintsev, A.
\newblock The building blocks of interpretability.
\newblock \emph{Distill}, 3\penalty0 (3):\penalty0 e10, 2018.

\bibitem[Prabhu \& Birhane(2020)Prabhu and Birhane]{prabhu2020large}
Prabhu, V.~U. and Birhane, A.
\newblock Large image datasets: A pyrrhic win for computer vision?
\newblock \emph{arXiv preprint arXiv:2006.16923}, 2020.

\bibitem[Raghu et~al.(2017)Raghu, Gilmer, Yosinski, and
  Sohl{-}Dickstein]{raghu2017svcca}
Raghu, M., Gilmer, J., Yosinski, J., and Sohl{-}Dickstein, J.
\newblock {SVCCA:} singular vector canonical correlation analysis for deep
  learning dynamics and interpretability.
\newblock In \emph{Advances in Neural Information Processing Systems 30: Annual
  Conference on Neural Information Processing Systems 2017, December 4-9, 2017,
  Long Beach, CA, {USA}}, pp.\  6076--6085, 2017.

\bibitem[Rauber et~al.(2016)Rauber, Fadel, Falcao, and
  Telea]{rauber2016visualizing}
Rauber, P.~E., Fadel, S.~G., Falcao, A.~X., and Telea, A.~C.
\newblock Visualizing the hidden activity of artificial neural networks.
\newblock \emph{IEEE transactions on visualization and computer graphics},
  23\penalty0 (1):\penalty0 101--110, 2016.

\bibitem[Roweis \& Saul(2000)Roweis and Saul]{roweis2000nonlinear}
Roweis, S.~T. and Saul, L.~K.
\newblock Nonlinear dimensionality reduction by locally linear embedding.
\newblock \emph{science}, 290\penalty0 (5500):\penalty0 2323--2326, 2000.

\bibitem[Selvaraju et~al.(2017)Selvaraju, Cogswell, Das, Vedantam, Parikh, and
  Batra]{selvaraju2017grad}
Selvaraju, R.~R., Cogswell, M., Das, A., Vedantam, R., Parikh, D., and Batra,
  D.
\newblock Grad-cam: Visual explanations from deep networks via gradient-based
  localization.
\newblock In \emph{{IEEE} International Conference on Computer Vision, {ICCV}
  2017, Venice, Italy, October 22-29, 2017}, pp.\  618--626. {IEEE} Computer
  Society, 2017.
\newblock \doi{10.1109/ICCV.2017.74}.

\bibitem[Springenberg et~al.(2014)Springenberg, Dosovitskiy, Brox, and
  Riedmiller]{springenberg2014striving}
Springenberg, J.~T., Dosovitskiy, A., Brox, T., and Riedmiller, M.
\newblock Striving for simplicity: The all convolutional net.
\newblock \emph{arXiv preprint arXiv:1412.6806}, 2014.

\bibitem[Sun et~al.(2014)Sun, Chen, Wang, and Tang]{sun2014deep}
Sun, Y., Chen, Y., Wang, X., and Tang, X.
\newblock Deep learning face representation by joint
  identification-verification.
\newblock In \emph{Advances in Neural Information Processing Systems 27: Annual
  Conference on Neural Information Processing Systems 2014, December 8-13 2014,
  Montreal, Quebec, Canada}, pp.\  1988--1996, 2014.

\bibitem[Szegedy et~al.(2015)Szegedy, Liu, Jia, Sermanet, Reed, Anguelov,
  Erhan, Vanhoucke, and Rabinovich]{szegedy2015going}
Szegedy, C., Liu, W., Jia, Y., Sermanet, P., Reed, S.~E., Anguelov, D., Erhan,
  D., Vanhoucke, V., and Rabinovich, A.
\newblock Going deeper with convolutions.
\newblock In \emph{{IEEE} Conference on Computer Vision and Pattern
  Recognition, {CVPR} 2015, Boston, MA, USA, June 7-12, 2015}, pp.\  1--9.
  {IEEE} Computer Society, 2015.
\newblock \doi{10.1109/CVPR.2015.7298594}.

\bibitem[Tenenbaum et~al.(2000)Tenenbaum, De~Silva, and
  Langford]{tenenbaum2000global}
Tenenbaum, J.~B., De~Silva, V., and Langford, J.~C.
\newblock A global geometric framework for nonlinear dimensionality reduction.
\newblock \emph{science}, 290\penalty0 (5500):\penalty0 2319--2323, 2000.

\bibitem[Van~der Maaten \& Hinton(2008)Van~der Maaten and
  Hinton]{van2008visualizing}
Van~der Maaten, L. and Hinton, G.
\newblock Visualizing data using t-sne.
\newblock \emph{Journal of machine learning research}, 9\penalty0 (11), 2008.

\bibitem[Wold et~al.(1987)Wold, Esbensen, and Geladi]{wold1987principal}
Wold, S., Esbensen, K., and Geladi, P.
\newblock Principal component analysis.
\newblock \emph{Chemometrics and intelligent laboratory systems}, 2\penalty0
  (1-3):\penalty0 37--52, 1987.

\bibitem[Zeiler \& Fergus(2014)Zeiler and Fergus]{zeiler2014visualizing}
Zeiler, M.~D. and Fergus, R.
\newblock Visualizing and understanding convolutional networks.
\newblock In \emph{European conference on computer vision}, pp.\  818--833.
  Springer, 2014.

\end{thebibliography}

\end{document}


\appendix

\section{Useful Links}
\textbf{Demo:} \url{https://umap-tour.github.io/index.html}\\
\textbf{Video:} \url{https://www.dropbox.com/sh/2za8rt3tpz35w7u/AAAKDpvgKDKvjRcthSTtexbja}\\
\textbf{Source code:} \url{https://github.com/umap-tour/source}\\

\section{Detail of Experiments}

\subsection{Computing Infrastructure}
All computations (the training and forward pass of neural networks, computation of UMAP embeddings) are conducted on a server with 12-core Intel(R) Xeon(R) CPU (E5-2603 v3 @ 1.60GHz) with 264 GB of memory. 
We also used one Tesla K80 GPU (with 11 GB memory) in the training and forward passing steps. 

\subsection{Detail of Training}
We visualized two models of GoogLeNet: one pre-trained model provided by PyTorch~\footnote{\url{https://umap-tour.github.io/googlenet.html}} and one trained on the above infrastructure~\footnote{\url{https://umap-tour.github.io/googlenet-train.html}}.
We trained GoogLeNet over the training set of ImageNet with simple image preprocessing (resize, central crop and normalize).
We optimize the cross-entropy loss (on both the main and auxiliary classifiers) with an Adam~\cite{kingma2014adam} optimizer using mini-batches of size 64.
The training takes 30 epochs ($\sim 60$ hours) to converge.

On both models, we observed cluster of humans, orientation of animals and animal faces.
On the model we trained, we observed the migration of dog-with-human images from the cluster of humans to the cluster of dogs from the 12-Inception to the 19-Linear layer~\footnote{\url{https://www.dropbox.com/s/zbxadke4gr6r479/2-layer-dynamics-dogs.mov}}.

\begin{figure}[th]
\centering
  \begin{subfigure}[t]{0.6\columnwidth}
    \centering
    \includegraphics[width=\columnwidth]{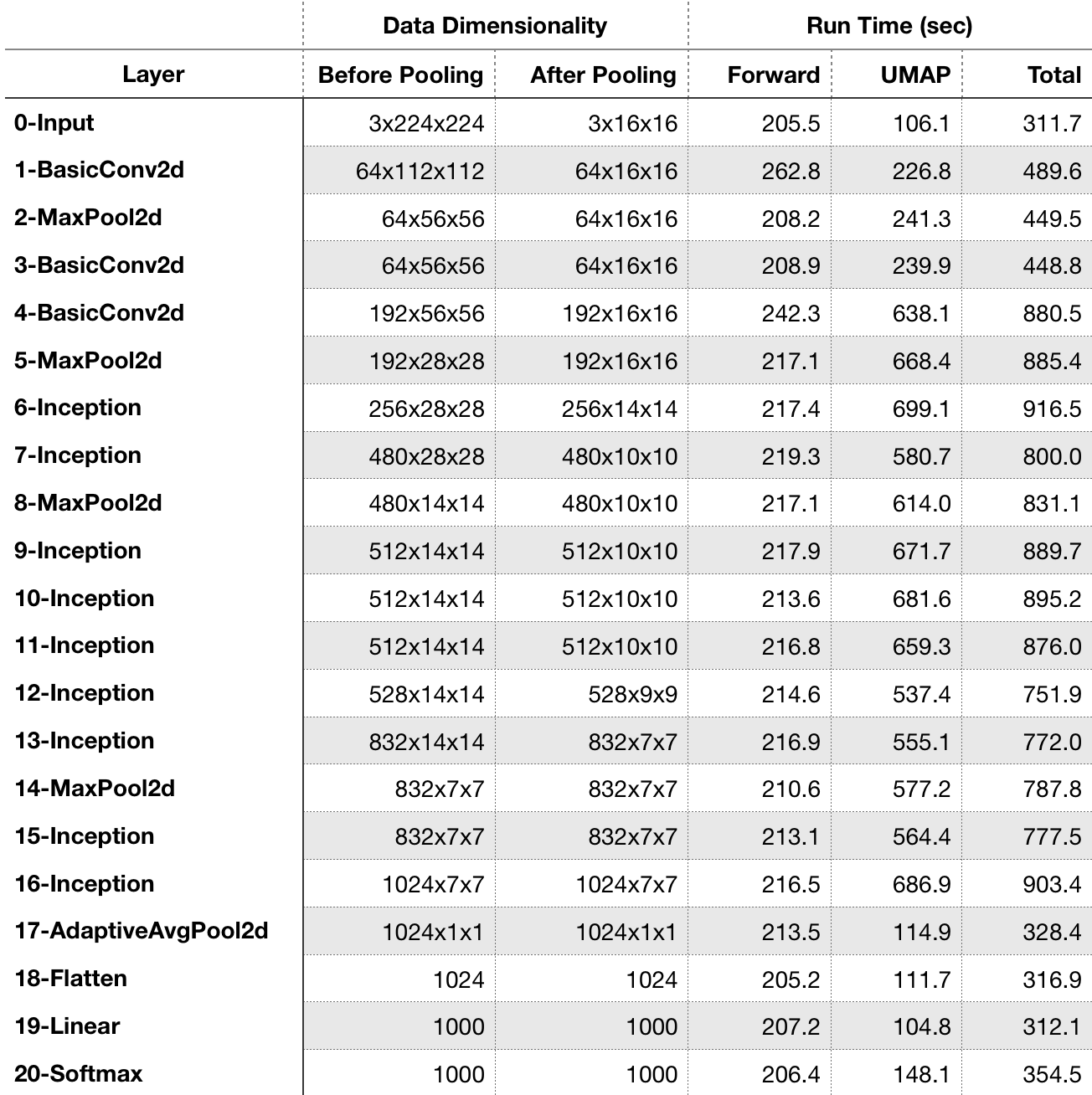}
    \newsubcap{
      Runtime of UMAP Tour when passing 50,000 ImageNet examples through GoogLeNet.
    }
    \label{table:runtime}
  \end{subfigure}
  \hfill
  \begin{subfigure}[t]{0.39\columnwidth}
    \centering
    \includegraphics[width=\columnwidth]{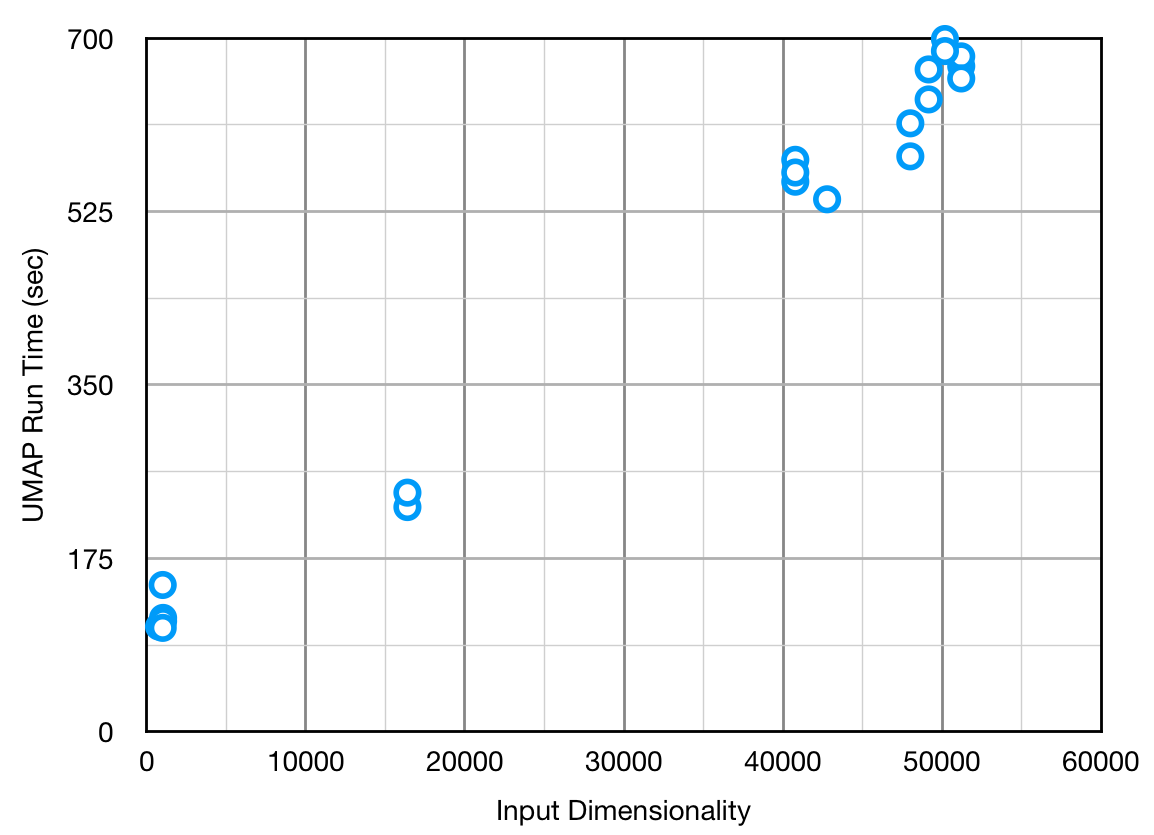}
    \newsubcap{
      UMAP's input dimensionality vs run time. The run time of UMAP embedding scales almost linearly with its input dimensions.
    }
    \label{fig:runtime-chart}
  \end{subfigure}
\end{figure}

\subsection{Runtime Complexity}
When processing data for UMAP Tour, we \circone pass image examples through the network and downsample the neuron activation through average pooling, \circtwo find the UMAP embedding of each layer \circthree compute pairwise similarity index among layers and \circfour visualize through the Grand Tour.
Here we will discuss the runtime for each of the steps.

The forward pass and downsampling together take no more than the forward pass of a neural network with poolings branching out of each layer of interest.
Although computing forward pass is dominated by the largest matrix multiplication among layers which takes up to $O(n^3)$ by naive implementations, its highly parallelizable nature makes it perform well on GPUs. 
Empirically, for a fixed number of examples (in our case, 50,000 validation examples of ImageNet), we observed near-constant runtime from layer to layer (Figure~\ref{table:runtime}) and linear growth with the number of examples.

The UMAP takes two major steps.
First, it estimates a kNN graph of the original space. 
Then, regarding the kNN graph as the ground truth, UMAP performs 'neighbor graph matching' on low dimensional embedding via stochastic gradient descent (SGD).
The first step is only performed once and nearest neighbor descent (NN-Descent)~\cite{heavner2019efficient} is used to estimate the kNN graph with high accuracy.
The author of NN-Descent reported an empirical cost of $O(n^{1.14})$ for all datasets they considered.
The second step, finding the embedding via (SGD) is performed iteratively until convergence.
Although the convergence rate of this step depends on data, overall, the UMAP reported a near-linear runtime complexity~\cite{mcinnes2018performance}.
The runtime for processing each layer of GoogLeNet are listed in Figure~\ref{table:runtime}. 
In Figure~\ref{fig:runtime-chart}, we plot run time of UMAP embedding as a function of its input dimensionality. 
The runtime is based on the passing 50000 ImageNet validation images to GoogLeNet and fitting the UMAP embeddings. 

Computing similarity index between layers through orthogonal Procrustes requires finding nuclear norms of $X^TY$, $X^TX$ and $Y^TY$. 
We used NumPy\cite{numpy} to compute these nuclear norms.
In NumPy, the nuclear norm is computed by finding and summing all singular values, which generally requires a cost of $O(mn^2)$ for computing the SVD of a $m \times n$ matrix.
However, we reduced the dimension of data with UMAP.
In UMAP Tour, computing similarity index from two UMAP embeddings, $X, Y \in \mathbb{R}^{n \times p}$, only requires matrix multiplications ($O(np^2)$) and SVD of very small matrices ($O(p^3)$, where $p=15$).

Finally, visualizing through the Grand Tour requires defining (a sequence of) projection matrices, projecting data points and matching layer representations, all in real time.
To achieve this, we lift some of the heavy computations to the GPU using WebGL.
Defining the Grand Tour projection (see the torus method in \cite{asimov1985grand}) at each time step $t$ requires $p(p-1)/2$ updates on the entries of the projection matrix.
In this step, we also post-process the projection matrix according to user's direct manipulation on data points.
Since $p$ is small ($p=15$) in our case, we implement this step on CPU (i.e. in JavaScript) and update the matrix as a uniform variable in GLSL.
Projecting data points to 2 or 3D requires matrix multiplications of complexity $O(np)$, we put this computation to the GPU (i.e. via GLSL) due to is highly parallelizable nature.
Finally, matching layer representations requires finding and applying an additional linear transform on one of the layers.
This part involves finding the SVD of $X^TY$, which takes $O(np^2)$ on CPU. 
To accelerate this part, we subsample a fraction (e.g. 0.1) of the $n$ examples and only align these examples.  
Putting all these together, we are able to use WebGL to render up to $50,000$ data points of $15$ dimensions at 60 fps on a 2560 x 1600 display.\footnote{See the demos in: \url{https://umap-tour.github.io/index.html}}

\begin{figure}[t]
\centering
\includegraphics[width=0.8\columnwidth]{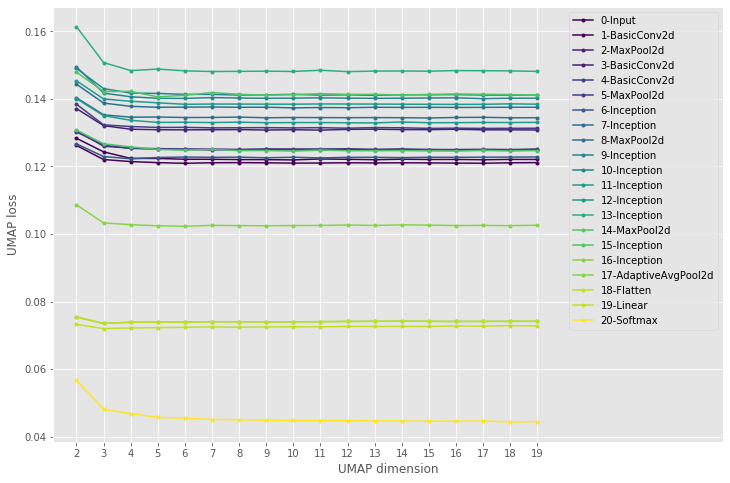}
\caption{
Analysis of UMAP embedding dimensions. 
UMAP's embedding loss depends on both the complexity of data and the dimensionality of the embedding. 
In general, later layers tend to have a lower UMAP loss. 
The UMAP loss decreases with the increase of embedding dimensions. 
In other words, for most layers the embedding quality benefits from having more than 3 dimensions.
}
\label{fig:umap-dim-analysis}
\end{figure}

\subsection{Impact of UMAP Embedding Dimensions}
We analyzed the impact of embedding dimension on the quality of UMAP.
For each layer we embed its neuron activations in different dimensions through UMAP and evaluate its embedding loss.
UMAP is a randomized algorithm but we only considered one embedding for each case, which already gives us enough information on the general trend.
In practice, one can run UMAP multiple times on the same data and pickup the one that gives the best quality. 
Figure~\ref{fig:umap-dim-analysis} shows the embedding loss on every layer of GoogLeNet.
In general, the embedding loss depends on both the complexity of data (i.e. layers) and the dimensionality of the embedding. 
Later layers typically to have a lower UMAP loss.

\section{More Examples and Use Cases}

\begin{figure*}[t]
\centering
  \begin{subfigure}[b]{0.8\columnwidth}
    \centering
    \includegraphics[width=\columnwidth]{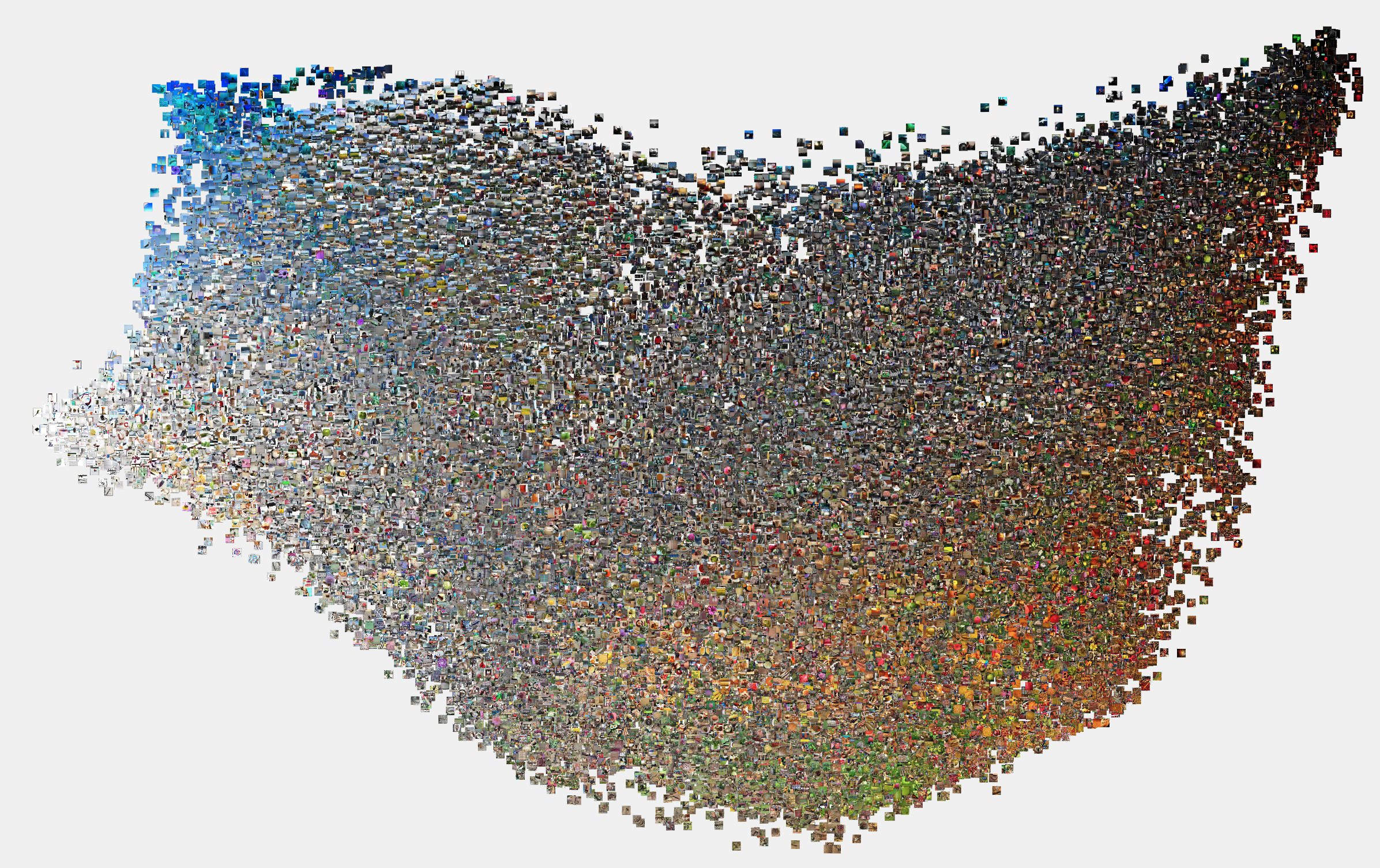}
    \newsubcap{
    On the 4-BasicConv2d layer, images are mostly arranged by colors.
    }
    \label{fig:4-BasicConv2d}
  \end{subfigure}
  \begin{subfigure}[b]{0.8\columnwidth}
    \centering
    \includegraphics[width=\columnwidth]{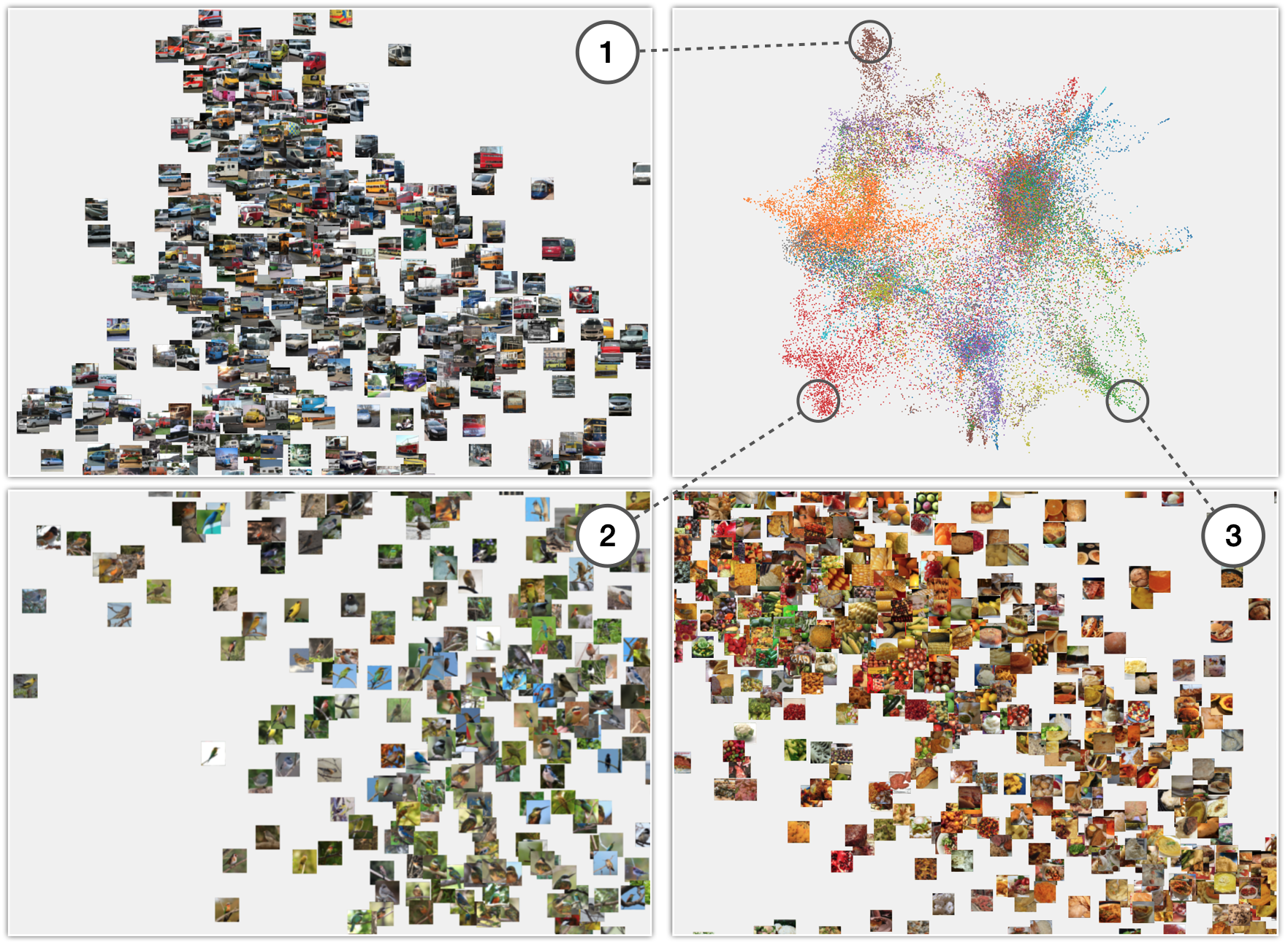}
    \newsubcap{
      Clusters found in the 15-Inception layer: \ding{192} cars, \ding{193} birds and \ding{194} food.
    }
    \label{fig:15-Inception}
  \end{subfigure}
\end{figure*}

\begin{figure}[t]
  \centering
  \includegraphics[width=\columnwidth]{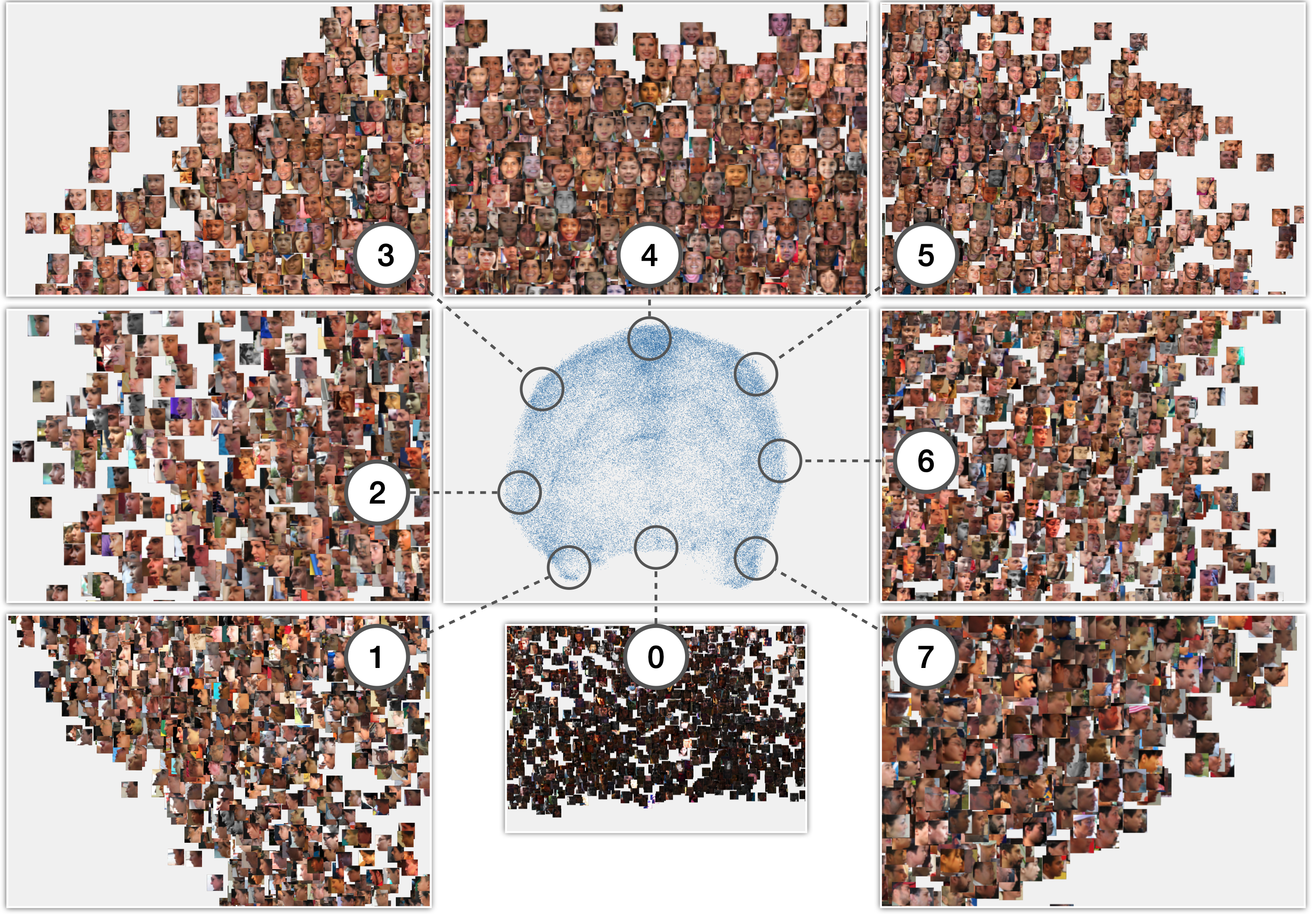}
  \caption{
    Orientation of faces (\ding{192} - \ding{198}: left to right) found in the 13-Inception layer.
  }
  \label{fig:13-Inception}
\end{figure}

Figure~\ref{fig:4-BasicConv2d}, \ref{fig:13-Inception} and \ref{fig:15-Inception} show other patterns we found within GoogLeNet when probing with the ImageNet or FairFace datasets.
Looking at ImageNet examples in the 4-BasicConv2d layer, we see images mostly arranged by colors (Figure~\ref{fig:4-BasicConv2d}).
On the 15-Inception layer, we observed clusters of cars, birds and food (Figure~\ref{fig:15-Inception}).
Pushing FairFace images to the 13-Inception layer, we found orientation of faces (Figure~\ref{fig:13-Inception}).

\begin{figure}[t]
\centering
\includegraphics[width=\linewidth]{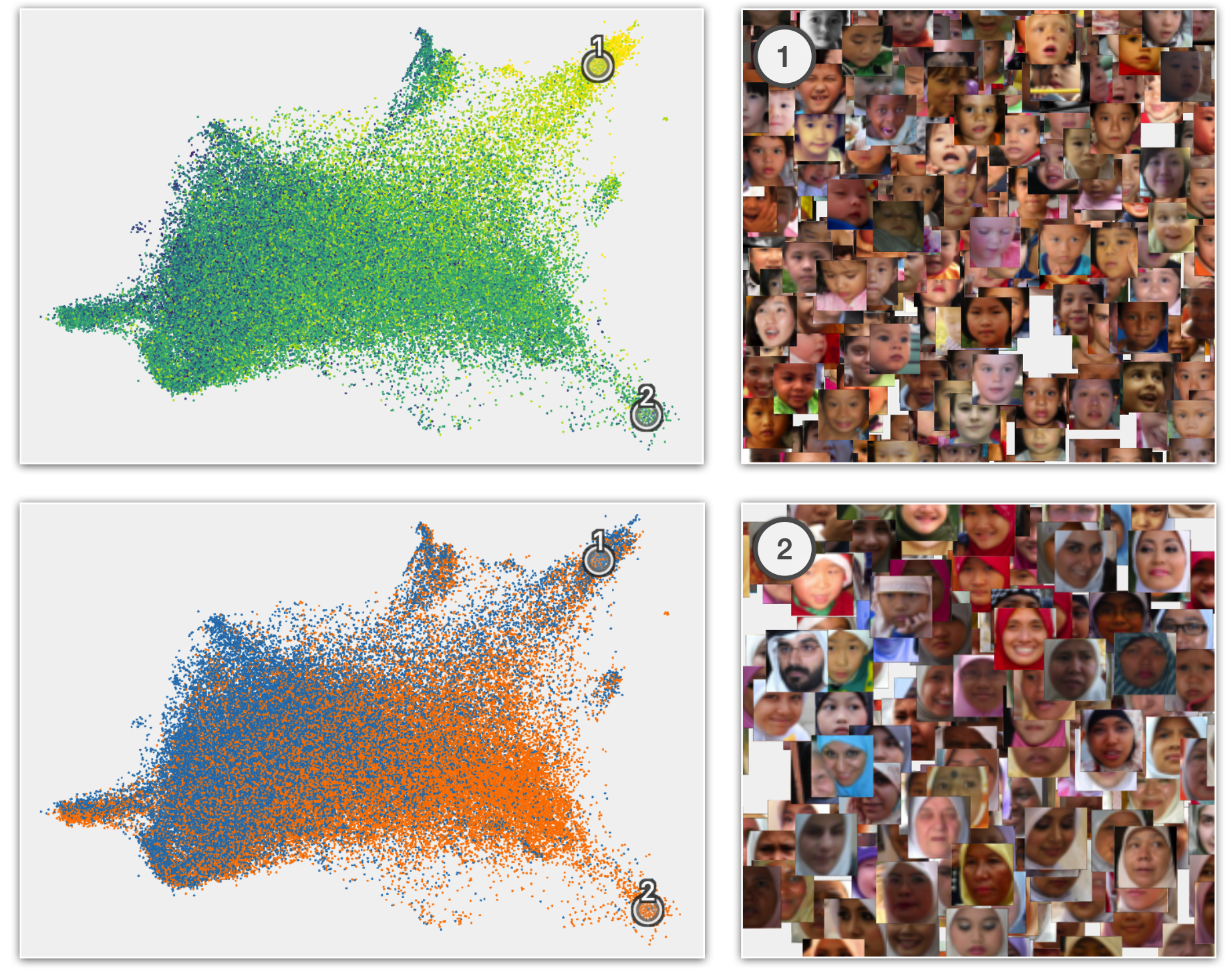}
\caption{
  Coloring data points by attributes from FairFace dataset reveals GoogLeNet's ability to identify certain age and gender groups.
  In the 16-Inception layer, 
  \textbf{top:} coloring by age (from \circyoung age 0-2 to \circold {\color{old}more than 70}) shows 0-2 age group in one cluster; 
  \textbf{bottom:} coloring by gender (\circmale {\color{male}males} \circfemale {\color{female}females}) exhibits veiled women in another cluster.
}
\label{fig:result-fairface}
\end{figure}

\subsection{Probing GoogLeNet with FairFace Dataset}
In the main section we found a human detector in GoogLeNet on the dataset that the model was trained on. 
We are curious about how well this human detector works, such as what patterns it captures, whether it biases toward certain race, gender, or age groups.
Since ImageNet do not have these attributes associated with the images, we used an external human face dataset, namely the FairFace dataset~\cite{karkkainen2019fairface}, for a deeper inspection.

FairFace contains human faces of various orientations, surroundings and lighting conditions.
For example some people wear sunglasses or stand behind microphones, and we can see through UMAP Tour that GoogLeNet is able to recognize these apparels or devices.
In addition, we utilized three attributes labeled for the FairFace examples: race, gender and age groups.
In UMAP Tour, we color data points by these attributes to see if any activation pattern is related to these attributes.
We found that GoogLeNet uses face to help recognize certain apparel or devices, and sometimes apparel happend to be related to certain age or race groups.
For example, in 16-Inception layer, GoogLeNet identifies baby faces and woman in veils, as shown in Figure~\ref{fig:result-fairface}.
This is not very surprising given the distinctive look of these two groups.
For most pictures of veiled women, GoogLeNet recognized the surrounded clothings as neck brace, abaya or bonnet, etc.
Similarly, baby faces have been classified by the look of their surroundings, recognized as bonnet, bib or bath towel, etc.

On the other hand, we did not find any global-scale pattern that is related to certain racial or ethnic groups.
This is not very surprising because GoogLeNet was not trained on face related tasks and FairFace well-balanced on every ethnic groups.
However, since certain apparel such as veil is closely related to religious customs and a face dataset typically does \textit{not} sample uniformly across race/ethnic groups \textit{conditioned} on these apparel, a globally well-balanced dataset might not distribute evenly in some local neighborhood in GoogLeNet internal layers.
For example, a large proportion of veils are worn by women from Southeast Asian or Middle Eastern, while greater proportion of recognized sunglasses are worn by white people.
These observations may certify concerns around reckless claims of pre-trained network in transfer learning.
For example, when someone claims to be able to recognize religious belief of people by fine tuning GoogLeNet, the network might simply take spurious features, such as veils, as signals.
In such cases, UMAP Tour is able to point out the problem with a direct description of the neural network internals.
%

\subsection{Comparing UMAP with PCA}

\begin{figure}[th]
\centering
\includegraphics[width=\linewidth]{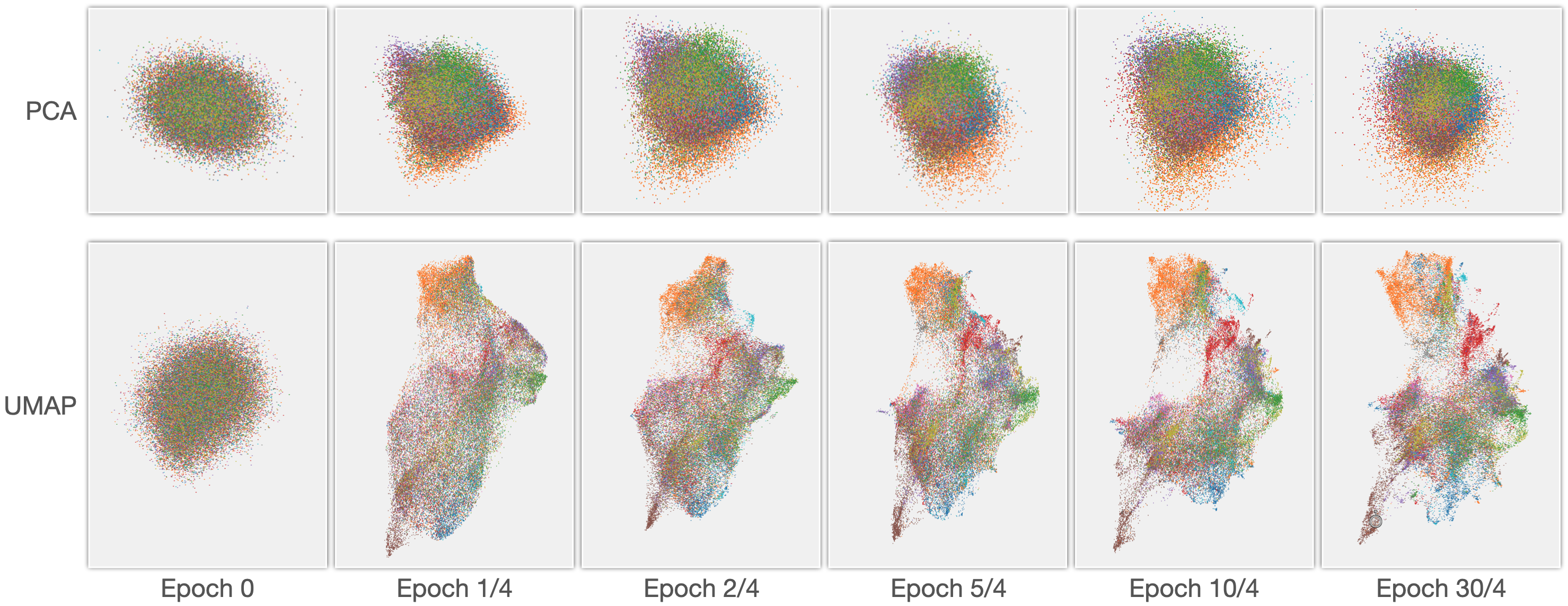}
\caption{PCA vs UMAP. PCA fails to reveal the refining steps in epoch 1/4 through 30/4.}
\label{fig:pca-training-dynamics}
\end{figure}

\subsubsection{Training Dynamics}
We compare the training dynamics given by UMAP to that given by principle component analysis (PCA).
As explained in the main text, we record the training once per $1/4$ epoch. 
For each recorded training stage, we find 15 principle directions via PCA and compare with the view given by UMAP.
Figure~\ref{fig:pca-training-dynamics} shows some sampled stages of training in a single layer of GoogLeNet and compare the dimensionality reduction plots given by PCA and UMAP.
We can see in the UMAP that from epoch 1/4 through 30/4, the overall configuration of examples remains the same, while the examples of each class get more and more concentrated, showing fine tuning steps in later epochs.
In PCA, although the look of overall configuration remains unchanged, it is much harder to see the fine tuning steps.
UMAP is able to reveal, sometimes exaggerate, the fine tuning steps due to its pulling force between examples and their k nearest neighbors - once the k nearest neighbors are getting closer and closer because of training, the pulling force imposed on the embedding gets larger and larger accordingly.
Note that the result in the visualization highly depends on the hyperparameters chosen for UMAP, and similar to t-SNE, absolute distances between data points or clusters might not mean anything~\cite{wattenberg2016use,coenen2019understanding}.

\begin{figure}[t]
\centering
\includegraphics[width=\linewidth]{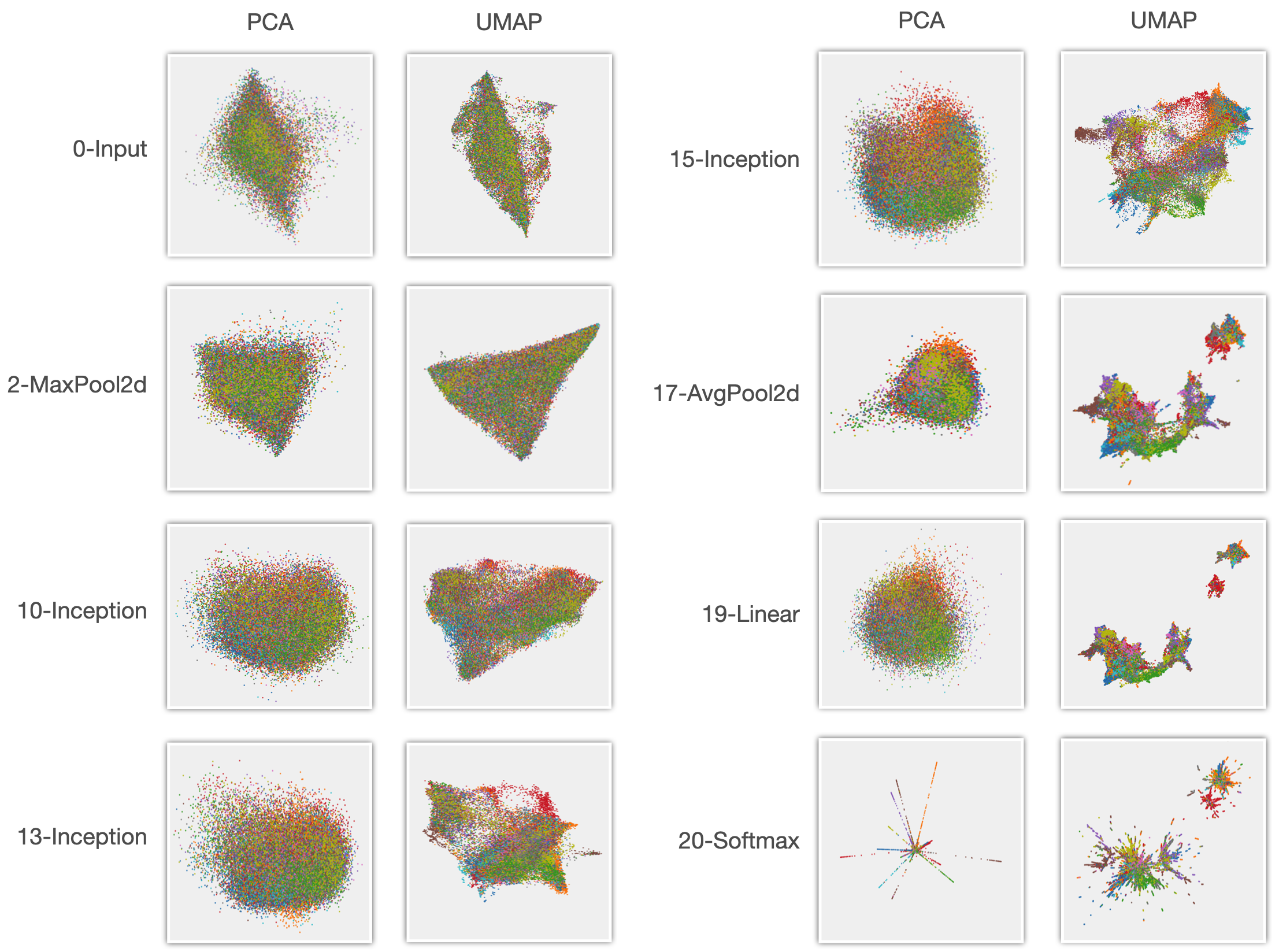}
\caption{PCA vs UMAP. PCA fails to capture interesting patterns within each layer.}
\label{fig:pca-layer-dynamics}
\end{figure}

\subsubsection{Layer Dynamics}
In Figure~\ref{fig:pca-layer-dynamics}, we compare the layer dynamics given by UMAP to that given by PCA.
Again, on each layer, we found 15 principle directions via PCA and compare with the view given by UMAP.
Although separations between classes can be found in later layers by filtering particular classes, we see 15-dimensional PCA fails to show finer differences between layers, compare to what is shown by UMAP.
Since most layers end with a batch normalization, the distribution of neuron activation vectors in these layers is almost spherical (i.e. having unit standard deviation in every direction along the standard basis), hence the variance in normalized data does not reliably reflect the importance of each direction.

\section{More Discussion}

\subsection{Orthogonal Procrustes or CKA?}
In main text, we highlighted connection between orthogonal Procrustes and CKA when they are used as similarity measures.
Here, we show their connection when they are used as \textit{alignments}.
We also highlight the strength of orthogonal Procrustes over CKA in visualizations.

\textbf{Alignment Induced by CKA}
Orthogonal Procrustes defines an alignment by finding the best rotation ($Q$) that matches one $p$-dimensional configuration $X$ to another $Y$ with the same dimension.
\begin{align*}
argmin_{Q}\; ||XQ - Y||_F^2\text{,\quad subject to }Q^T Q = I
\end{align*}
Similarly, centered kernel alignment (CKA) also comes with an alignment. 
As mentioned in \cite{kornblith2019similarity}, the alignment ($W$) induced by CKA can be formulated in a similar criterion:
\begin{align}
argmin_{W}\; ||XW - Y||_F^2\text{,\quad subject to } tr(W^T W) = 1
\label{eq:w}
\end{align}
Note this only differs from orthogonal Procrustes in the constraint.
The constraint of CKA is more flexible than orthogonal Procrustes, since in orthogonal Procrustes $Q^T Q = I \Rightarrow tr(Q^T Q) = p$, a scalar factor away from the constraint of CKA.
To better compare these two constraints, let us scale $W$ by $\sqrt{p}$:
\begin{align*}
\hat{W} = \sqrt{p} \cdot W
\end{align*}
so that
\begin{align*}
tr(\hat{W}^T \hat{W}) = p = tr(Q^T Q)
\end{align*}

Geometrically, the new constraint requires the (scaled) alignment $\hat{W}$ to 'preserve the diagonal length of the unit box'.
To see this, consider a unit box (with length $=1$ on each side, hence has diagonal length $=p^{1/2}$) living in the space of $X \in \mathbb{R}^p$.
Note $tr(\hat{W}^T \hat{W})$ is the sum of squared singular values of $\hat{W}$ and the trace equals to $p$,
\begin{align*}
p = tr(\hat{W}^T \hat{W}) = \sum_{i=1}^p \hat{\sigma}_i^2 
\end{align*}
In other words, transforming the unit box by $\hat{W}$ gives a box with side lengths $\hat{\sigma}_i$.
It keeps the L2 norm of the box's diagonal as a constant.
Moreover, the optimal alignment under the criterion (Eq.\ref{eq:w}) can be found via Lagrange multiplier.
\begin{align*}
W^* = \frac{1}{z} X^T Y
\end{align*}

\textbf{CKA in visualizations}
Comparing to the alignment given by orthogonal procrustes, $Q=UV^T$ where $U$ and $V^T$ comes from the SVD of $X^TY$, this alignment retains the scaling factors in $X^TY$.
To us, this is \textit{not} a desirable property for visualizing purposes.
When visualizing layer representations $X$ or $Y$ via scatter plots, one wants to display the data variance in the plot.
However, when switching from layer $Y$ to $X$, the alignment $W$ scales the target layer $X$ so that it aligns against $Y$, which hides away some of the variance of $X$ in favor of the variance in $Y$. 
Due to the scaling factors \textit{in the alignment}, we consider CKA alignments less suitable than orthogonal Procrustes in visualizations.


\subsection{Orthogonal Procrustes or SVD?}
Comparing to the SVD method used by Li et al.~\cite{li2020visualizing} which only work for consecutive (and linearly connected) layers, orthogonal Procrustes generates a slightly different alignment between linearly connected layers.
Recall that when two layers $X$ and $Y$ are related by a linear transformation $Y=XW$, Li et al. align two layers by the SVD of $W$ - i.e. they define the alignment to be $U_W V^T_W$ where $U_W$ and $V_W$ are left and right singular matrices of $W = U_W \Sigma_W V^T_W$.
Then they interpolate layers by $X(s) = X U_W((1-s) \cdot I + s \cdot \Sigma_W) V^T_W$.
In other words, their animated layer-by-layer transformations is explaining the scaling \textit{action} of $W$. 
Meanwhile, orthogonal Procrustes takes the SVD of $X^TY = X^TX W$, which takes an additional factor, $X^TX$, into account.
Compare to Li et al, the resultant interpolation of our method considers the variance in $X$ on top of $W$ (or covariance between $X$ and $Y$), which means that orthogonal Procrustes is explaining the \textit{difference} between the two data configurations.
The two approaches can collapse in some trivial cases - when $X$ is the identity matrix, the two alignments are exactly the same.
More generally, if $X$ varies uniformly along all directions, which is nearly the case under certain circumstances (e.g. after batch normalization layers), $X^TX$ would have singular values close to 1.
In that case the two methods tend to generate similar alignments.



\section{Limitations}

\subsection{Pooling only applies to CNNs?}
Recall that we used average pooling to reduce the dimensionality of neuron activations before computing their UMAP embeddings.
Note that pooling methods we used only applies to convolutional neural networks. 
Despite also being applied in other visualization systems\cite{hohman2019s}, currently we consider this as a limitation to our method. 
In the future, we would like to consider alternatives for other architectures, e.g. RNNs.

\subsection{Visualization only work best for images?}
Notice that in UMAP Tour, the image examples as shown as colored marks or thumbnails of pictures.
On other types of data, such as sound waves or sentences, the thumbnails might not be able to summarize the data.
We would have to consider better and compact representations of examples in fields other than visions. 


\newpage
\bibliographystyle{template/icml2020}
\bibliography{bib}